\documentclass{aastex701}

\usepackage{rotating}
\usepackage{graphicx}
\def\ctscm2s{cts\,cm$^{-2}$\,s$^{-1}$}

\usepackage{chngcntr}
\usepackage{amsmath}
\usepackage{url}
\usepackage{fixltx2e}
\usepackage{threeparttable}
\usepackage{graphicx}
\usepackage[english]{babel}
\usepackage{rotating,tabularx}
\usepackage{makecell}

\usepackage[T1]{fontenc}
\usepackage{bm}

\usepackage{array}
\usepackage{ulem}

\usepackage{subfigure}

\begin{document}

\counterwithout{figure}{section}
\renewcommand{\thefigure}{\arabic{figure}}

\title{Multi-mission Investigation of X-ray Superorbital Modulation in the Supergiant High Mass X-ray Binary 4U 1538--52}

\author[0009-0001-9108-7249]{Hale I. Cohen}
\affil{Institute for Astrophysics and Computational Sciences, Department of Physics, The Catholic University of America,
Washington, DC 20064, USA}
\email{cohenh@cua.edu}
\affiliation{Center for Space Sciences and Technology, University of Maryland, Baltimore County, 1000 Hilltop Circle, Baltimore, MD 21250, USA}

\author[0000-0002-2413-9301]{Nazma Islam}
\email{nislam@umbc.edu}
\affiliation{Center for Space Sciences and Technology, University of Maryland, Baltimore County, 1000 Hilltop Circle, Baltimore, MD 21250, USA}
\affiliation{X-ray Astrophysics Laboratory, Code 662 NASA Goddard Space Flight Center, Greenbelt Rd., MD 20771, USA}

\author[0000-0002-3396-651X]{Robin H.~D. Corbet}
\email{corbet@umbc.edu}
\affiliation{CRESST II  and X-ray Astrophysics Laboratory, Code 662 NASA Goddard Space Flight Center, Greenbelt Rd., MD 20771, USA}
\affiliation{Center for Space Sciences and Technology, University of Maryland, Baltimore County, 1000 Hilltop Circle, Baltimore, MD 21250, USA}
\affiliation{Maryland Institute College of Art, 1300 W Mt Royal Ave, Baltimore, MD 21217, USA}

\author[0000-0003-3540-2870]{Alexander Lange}
\email{alexlange@email.gwu.edu}
\affiliation{X-ray Astrophysics Laboratory, Code 662 NASA Goddard Space Flight Center, Greenbelt Rd., MD 20771, USA}
\affiliation{Center for Space Sciences and Technology, University of Maryland, Baltimore County, 1000 Hilltop Circle, Baltimore, MD 21250, USA}
\affiliation{Department of Physics, The George Washington University, Washington, DC 20052, USA}

\author[0000-0001-7532-8359]{Katja Pottschmidt}
\thanks{deceased 17 June 2025}
\email{katja@umbc.edu}
\affiliation{Center for Space Sciences and Technology, University of Maryland, Baltimore County, 1000 Hilltop Circle, Baltimore, MD 21250, USA}
\affiliation{Code 661 Astroparticle Physics Laboratory, NASA Goddard Space Flight Center, Greenbelt, MD 20771, USA}

\author[0000-0002-8201-1525]{Enrico Bozzo}
\email{Enrico.Bozzo@unige.ch}
\affiliation{Department of Astronomy, Universit\'e de Gen\`eve, 16 chemin d'\'Ecogia, 1290 Versoix, 
    Switzerland}
\affiliation{INAF -- Osservatorio Astronomico di Brera,  Via E.\ Bianchi 46, I-23807, Merate, Italy}

\author[0000-0003-0258-7469]{Patrizia Romano}
\email{patrizia.romano@inaf.it}
\affiliation{INAF -- Osservatorio Astronomico di Brera,  Via E.\ Bianchi 46, I-23807, Merate, Italy}

\author[0000-0002-1118-8470]{Ralf Ballhausen}
\email{ballhaus@umd.edu}
\affiliation{X-ray Astrophysics Laboratory, Code 662 NASA Goddard Space Flight Center, Greenbelt Rd., MD 20771, USA}
\affiliation{Department of Astronomy, University of Maryland, College Park, MD 20742, USA}

\author[0000-0002-0380-0041]{Christian Malacaria}
\email{cmalacaria.astro@gmail.com}
\affiliation{INAF -- Osservatorio Astronomico di Roma, Via Frascati 33, 00076, Monte Porzio Catone, (RM), Italy;}

\author[0000-0001-7532-8359]{Joel B. Coley}
\email{joel.coley@howard.edu}
\affiliation{CRESST II  and X-ray Astrophysics Laboratory, Code 662 NASA Goddard Space Flight Center, Greenbelt Rd., MD 20771, USA}
\affiliation{Department of Physics and Astronomy, Howard University, Washington, DC 20059, US}

\begin{abstract}

Superorbital modulations has been detected in the supergiant High-Mass X-ray binary 4U 1538–52 using long-term monitoring with the Neil Gehrels Swift Observatory Burst Alert Telescope (BAT). The source also exhibits a long-term pulse period evolution as seen with Rossi X-ray Timing Explorer (RXTE), INTEGRAL, and Fermi Gamma-Ray Burst Monitor (GBM) that appears uncorrelated with changes in its X-ray flux. To investigate the mechanisms causing these superorbital modulations and its possible dependence on pulse period changes, we analyzed long-term monitoring with Swift-BAT and Monitor of All Sky X-ray Image Gas Slit Camera (MAXI-GSC) to construct dynamic power spectra and superorbital intensity profiles. In addition, we used pointed X-ray observations from Nuclear Spectroscopic Telescope Array (NuSTAR) and Neutron Star Interior Composition Explorer mission (NICER) to investigate the pulsation and spectral properties across different superorbital and orbital phase intervals. We find the presence of superorbital modulations in the MAXI-GSC 2–20\,keV lightcurves, consistent with the periodicity observed with the Swift-BAT lightcurves. However, no significant changes are detected in the pulse profiles or spectral parameters across different superorbital, orbital, or pulse-change intervals. This lack of spectral or timing variations with orbital and superorbital phases suggests that the mechanisms driving the observed superorbital modulation and pulse period changes are likely associated with large-scale stellar wind structures, such as Co-Rotating Interaction regions, within the stellar wind of the supergiant companion.
\end{abstract}


\section{Introduction} \label{sec:intro}

High-Mass X-ray Binaries (HMXBs) consist of a compact object, generally a Neutron Star (NS) accreting from the stellar wind of a massive ($\geq$ 10 $M_\odot$) stellar companion.
HMXBs can be divided into Be High-Mass X-ray Binaries (BeHMXBs), which have Be-type stellar companions; Roche Lobe Overflow (RLO) systems, which have a companion which fills its Roche lobe and causes matter to flow out and form a persistent accretion disk; and supergiant High-Mass X-Ray Binaries (sgHMXBs), which have supergiant O and B type stellar companions. BeHXMBs accrete from a circumstellar `decretion' disk of material that forms around the Be companion star and RLOs accrete from their persistent accretion disks, while sgHMXBs accrete via the stellar wind. There are three types of periodicities that have been observed in HMXBs with an NS as the compact object: pulsations on the rotation period of the NS (which is a few seconds to about 1,000\,seconds for typical HMXBs), the orbital period of the binary system (typically a few days to hundreds of days) and a superorbital period (few tens to 100\,days) \citep{KotzeCharles2012, CorbetKrimm2013, corbet2021}. For HMXB systems like Her X--1 and SMC X--1 which contain an accretion disk, the superorbital modulations are unstable and are thought to be due to the precession of the accretion disk or the NS \citep{Ogilvie2001, Doroshenko2022, Clarkson2003, Trowbridge2007, Brumback2020}. For BeHMXBs, the superorbital modulations are related to the formation and dissipation of the circumstellar decretion disk around the Be star as well as one-armed oscillations of the disk \citep{Rajoelimanana2011}.
\par
sgHMXBs accrete matter primarily through the stellar wind, with a transient accretion disk present in few systems such as 2S 0114+650 \citep{Hu2017} and OAO 1657-415 \citep{Jenke2012}. There are six currently known sgHMXBs that display superorbital modulation. Listed in order of discovery, they are 2S 0114+650 \citep{Farrell2006}, IGR J16493--4348, IGR J16479--4514, 4U 1909+07, IGR J16418--4532 \citep{CorbetKrimm2013}, and 4U 1538–52 \citep{corbet2021}, as well as IGR J16393--4643 where superorbital modulation have been tentatively detected \citep{corbet2021}. Different theories have been proposed to explain these superorbital modulations in sgHMXBs, such as tidally induced oscillations in nonsynchronously rotating stars \citep{koenigsberger2006}, precessing transient accretion disks \citep{Hu2017}, and the presence of large-scale density and velocity structures in the stellar winds of the OB supergiant stars, called Corotating Interaction Regions \citep[CIRs; ][]{Bozzo2017}. 
\par
4U 1538--52 (sometimes referred to as 4U 1538--522) is a wind-accreting sgHMXB, which consists of a $1.3 \pm 0.2 M_\odot$ NS and a $19.9 \pm 3.4 M_\odot$ B0 Iab companion QV Nor \citep{Reynolds1992}. It was discovered in 1974 with the Uhuru satellite \citep{Giacconi1974} and has an orbital period of about $\sim$3.73\,d and exhibits regular eclipses persisting for for $\sim$0.6 d \citep{Hemphill2019, Davison1977}. The distance to the binary is estimated to be 6.82$\pm$0.02\,kpc from Gaia Data Release 3 \citep{bailer2021}. Superorbital modulations have been detected in long-term Swift-BAT lightcurves at the superorbital period, whose value was previously calculated to be $14.9130 \pm 0.002$6\,d \citep{corbet2021}.
A Cyclotron Resonance Scattering Feature (CRSF or `cyclotron line') is present in the X-ray spectrum at around 22\,keV, first observed by \textit{GINGA} \citep{Clark1990}, and subsequently by \textit{Rossi X-ray Timing Explorer} \citep[RXTE, ][]{Coburn2001,Rodes-Roca2009}, INTErnational Gamma-Ray Astrophysics Laboratory \citep[INTEGRAL, ][]{Hemphill2013}, Suzaku \citep{Hemphill2014}, Nuclear Spectroscopic Telescope Array   \citep[NuSTAR, ][]{Hemphill2019} and AstroSAT  \citep{Varun2019}. 
\par
The long-term pulse period history of 4U 1538--52 shows episodes of at least three torque reversals from MJD 50450 to MJD 60741: spin-down to spin-up in 1990 \citep{Bildsten1997,Rubin1997}, spin-up to spin-down around 2009 \citep{Hemphill2013}, and a sinusoidal variation in the pulse period since 2019 \citep{Sharma2023, Hu2024}. These torque reversals are not found to be associated with any changes in the X-ray flux. The origin of these pulse-period changes and torque reversals have been attributed to long-term changes in the stellar wind, like those seen in Vela X--1 \citep{liao2022, kretschmar2021, Sharma2023, Hu2024, ElMellah2019}, the presence of transient accretion disks \citep{Hu2017}, or tidal interactions between the massive companion and the NS \citep{kim2025}. \cite{corbet2021} suggested a connection between the torque reversal and the superorbital modulations in 4U 1538--52, similar to that seen in 2S 0114+650, another sgHMXB showing correlated changes in the superorbital modulations with the pulse period changes \citep{Hu2017}. 
\par
In this manuscript, we investigate the connection between the superorbital modulations and the pulsation and spectral characteristics as well as changes in the pulse period of 4U 1538--52, using long-term observations of 4U 1538--52 with Fermi-GBM, Swift-BAT, and MAXI-GSC and several pointed X-ray observations with NICER and NuSTAR. In Section~\ref{sec:Long-Term Observations}, we study the long-term X-ray flux and pulse period behavior using Fermi-GBM, Swift-BAT, and MAXI-GSC monitoring. The timescales and changes in the superorbital modulations are investigated by constructing dynamic power spectrum using Swift-BAT and MAXI-GSC lightcurves in Section \ref{sec:dynamic power + intensity profiles}. In Section~\ref{sec:Pointed X-ray Observations}, we investigate the pulsation and spectral characteristics of several pointed NICER and NuSTAR observations as well as several X-ray archival observations, as a function of the orbital and superorbital phase. In Section \ref{sec:Results} we discuss these results and we discuss them in the context of models used to explain the presence of superorbital modulations, pulse period changes, and torque reversals in sgHMXBs.

\begin{figure}
  \centering
  \includegraphics[width=.75\textwidth]{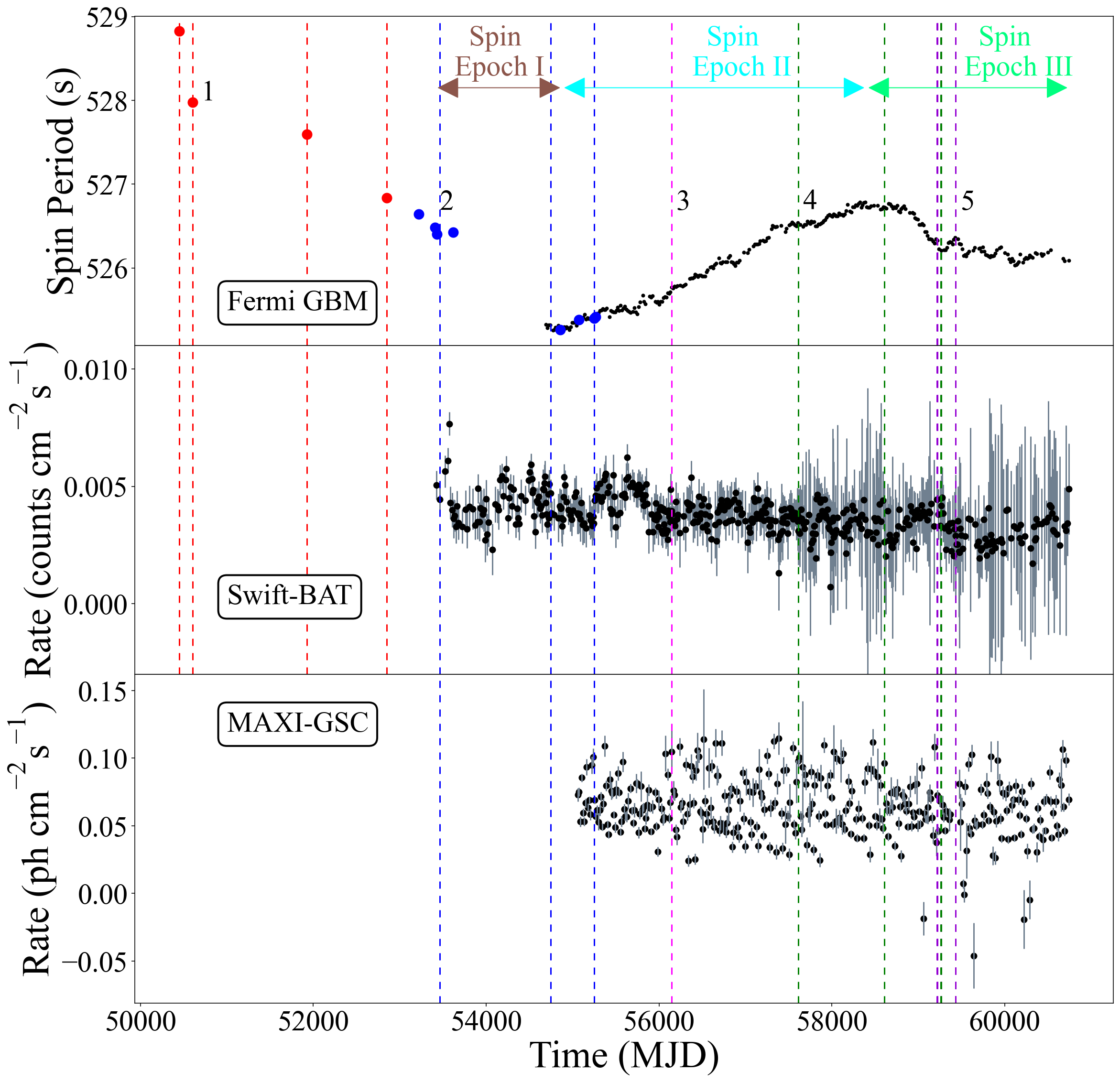}
  \caption{Top: Spin-period of 4U 1538--52 from 15 years of Fermi-GBM observations (black points) along with pulse period measurements from RXTE (red points), and INTEGRAL (blue points) observations. Dashed lines indicate dates of pointed X-ray observations of the source: RXTE (red), INTEGRAL (blue), Suzaku (pink), NuSTAR (green) and NICER (purple). Epochs of pulse period changes are marked with arrows; Spin Epoch I (brown), Spin Epoch II (cyan), and Spin Epoch III (light green). Middle: The long-term lightcurves of Swift-BAT (15--50\,keV) observations. Bottom: long-term lightcurves of MAXI-GSC (2--20\,keV). The Swift-BAT and MAXI-GSC lightcurves have been re-binned with a time period of 14.9081\,d (superorbital period)}
  \label{fig:4u1538 fermi}
  \end{figure}

\section{Data and Analysis}
\label{sec:Data and Analysis}
\subsection{Fermi-GBM, Swift-BAT and MAXI-GSC Long-Term Observations}
\label{sec:Long-Term Observations}
The Fermi Gamma-ray Space Telescope operates the Gamma-Ray Burst Monitor \citep[GBM;][]{Meegan2009}. In the top panel of Figure~\ref{fig:4u1538 fermi}, we plot the pulse frequency history of 4U 1538--52 which is available through the Fermi-GBM Accreting Pulsar Program   \citep[GAPP\footnote{\href{https://gammaray.nsstc.nasa.gov/gbm/science/pulsars/lightcurves/4u1538.html}{https://gammaray.nsstc.nasa.gov/gbm/science/pulsars/lightcurves/4u1538.html}},][]{Finger2009, Malacaria2020} and are corrected for the binary orbit. We also mark the dates of the observations carried out by the RXTE, INTEGRAL, Suzaku, NuSTAR and NICER, which are tabulated in Table \ref{Tab:observation_list}. The pulse period changes can be divided into three epochs: Spin Epoch I from MJD 53415 (February, 2005) to 54880 (February, 2009) when the pulsar was in a spin-up phase with torque reversal around MJD 54880, Spin Epoch II from MJD 54880 to MJD 58400 (October, 2018) when the pulsar was in a spin-down phase with a torque reversal around MJD 58400, and Spin Epoch III from MJD 58400 to MJD 60750 (March, 2025) with an erratic sinusoidal spin change with an spin-up trend \citep{Sharma2023}. The spin-up phase from 2005 to 2009 was also observed with CGRO-BATSE, which is not used in this work \citep{Rubin1997}.
\par
Swift-BAT \citep{Barthelmy2005:BAT} is a hard X-ray telescope with a coded aperture mask that operates in the 14--195\,keV energy band. We used orbital lightcurves from the BAT Transient Monitor\footnote{\href{https://swift.gsfc.nasa.gov/results/transients/}{https://swift.gsfc.nasa.gov/results/transients/}} \citep{krimm2013swift} in the 15--50\,keV energy band from MJD 53415 to MJD 60750. The lightcurves were further screened to exclude bad quality points and to only use the data for which the data-quality flag “DATA$\_$FLAG” was set to 0. A small number of data points with very low fluxes and unrealistically small uncertainties were also removed from the lightcurves \citep{CorbetKrimm2013}. The middle panel of Figure \ref{fig:4u1538 fermi} shows the Swift-BAT lightcurve folded and binned on the revised superorbital period of 14.908\,d obtained in Section \ref{sec:dynamic power + intensity profiles}.
\par
MAXI \citep{MAXI:Matsuoka2009PASJ, Mihara2011} is an all-sky X-ray monitor which consists of two detectors, the Solid-state Slit Camera (SSC) and the Gas Slit Camera (GSC) operating on the International Space Station (ISS). We use the MAXI-GSC lightcurves in 2--20\,keV energy band from MJD 55053 to MJD 60751 \footnote{\href{https://maxi.riken.jp/star\_data/J1542-523/J1542-523.html}{https://maxi.riken.jp/star\_data/J1542-523/J1542-523.html}}. We filtered some unusually low count-rates in the MAXI-GSC lightcurves, which are unphysical in origin, and plotted the cleaned MAXI-GSC lightcurve, binned with the revised superorbital period of 14.9081\,d, in the bottom panel of Figure~\ref{fig:4u1538 fermi}. The Swift-BAT and MAXI-GSC lightcurves were corrected to the solar system barycenter using {\tt astropy time} package\footnote{\href{https://docs.astropy.org/en/stable/time/}{https://docs.astropy.org/en/stable/time/}}.
\par
\subsubsection{Dynamic Power Spectra and Superorbital Intensity Profiles}
\label{sec:dynamic power + intensity profiles}
\begin{figure}
\centering
    \includegraphics[width=0.65\linewidth]{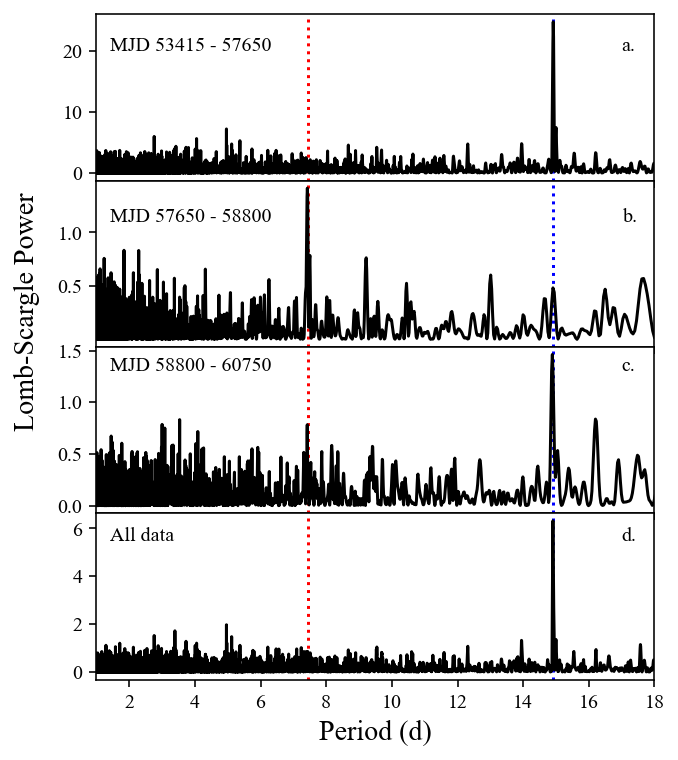}
\caption{Power spectra of Swift-BAT data calculated using Lomb-Scargle periodograms. The fundamental frequency of the superorbital period (14.9081 d) is marked by the dotted blue line, and the second harmonic of the superorbital period (7.4541 d) is marked by the dotted red line. a: The power spectra of the time interval MJD 53415 till 57650. b: Power spectra of the time interval MJD 57650 till 58800. c: Power spectra of the time interval MJD 58800 till 60750. d: Power spectra of the entire observation period.} 
\label{fig:power_spectra}
\end{figure}

\begin{figure}
\centering
    \includegraphics[width=0.95\linewidth]{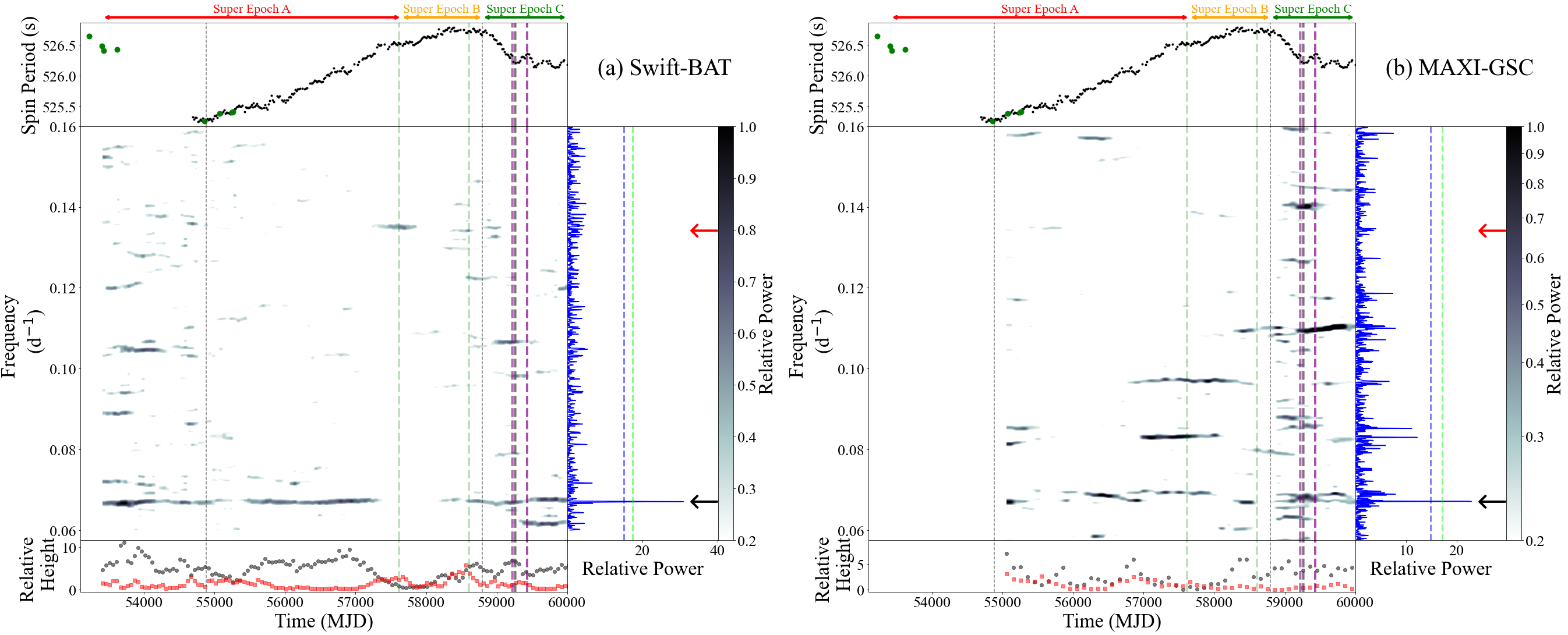}
\caption{Dynamic power spectra of Swift-BAT (a) and MAXI-GSC (b) data. The dates of pointed X-ray observations of 4U 1538--52 (see Table \ref{Tab:observation_list}) are marked by green (NuSTAR) and purple (NICER) dashed lines. The thin black lines represent the different epochs of pulse period changes as mentioned in Section~\ref{sec:Long-Term Observations}. Top: The pulse period of 4U 1538--52 from Fermi-GBM (black) and INTEGRAL (green) observations are plotted here. Middle: Dynamic power spectra constructed using LSP.
Right: Power spectrum of the lightcurve, normalized to the average power. The fundamental frequency of the superorbital period is marked by a black arrow and the second harmonic by a red arrow, with 99.9\% and 99.99\%
significance levels indicated by the dashed blue and green lines respectively. 
Bottom: The height of the fundamental frequency of the superorbital modulations (black) and the second harmonic (red) relative to the mean power.}
\label{fig:dynamic_power_spectra}
\end{figure}

To monitor the long-term changes in the strength of the superorbital modulations in 4U 1538--52, we constructed power spectra using Lomb-Scargle Periodogram \citep[LSP, ][]{lomb1976, scargle1982, vanderplas2018} of the Swift-BAT and MAXI-GSC lightcurves. We filtered out the times that correspond to the phases of the orbital eclipse from ($0.85<\phi_{\mathrm{orbital}}<1.00$ and $0.00<\phi_{\mathrm{orbital}}<0.15$) for the Swift-BAT lightcurves and ($0.6<\phi_{\mathrm{orbital}}<1.00$ and $0.00<\phi_{\mathrm{orbital}}<0.3$) for MAXI-GSC, using time of the mid-eclipse T$_{\pi/2}$ = MJD 57612.53 and an orbital period of 3.72831\,d \citep{Hemphill2019}, a value which we used in subsequent analysis. The power spectrum was calculated using the semi-weighting technique, in which the error bar on each data point and the excess variability in the light curve was taken into account as described in \cite{corbet2007}. In Figure \ref{fig:power_spectra}, we show the Swift--BAT power spectra for the different time intervals where the fundamental frequency and harmonic were present, also presented in \cite{corbet2021}, and the extended time interval MJD 58800 till 60750. To create the dynamic power spectra, the power spectrum of the LSP was calculated using 750\,d time spans that were successively shifted in time by 50\,d relative to each other, using the semi-weighting technique described above. The top panels of Figure \ref{fig:dynamic_power_spectra}a and b show the pulse period evolution (same as top panel of Figure \ref{fig:4u1538 fermi}), and the middle panels show the dynamic power spectrum constructed using 15--50\,keV Swift-BAT and 2--20\,keV MAXI-GSC lightcurves. The bottom panels show the relative height of fundamental frequency (and harmonic for Swift-BAT dynamic power spectra), which is calculated by taking the power per bin and dividing by the mean power. We estimate a revised superorbital period of 14.9081$\pm$0.0019\,d from the BAT and GSC power spectra for the entire duration of the lightcurves (right panels of Figure \ref{fig:dynamic_power_spectra}a and b). This is within the 2$\sigma$ error-bars of the previously superorbital period of 14.9130$\pm$0.0026\,d found by \cite{corbet2021} using Swift-BAT data from MJD 53416 till MJD 58880. 
The fundamental frequency of the superorbital modulations is marked by a black arrow, and it is not always strongly present in the Swift--BAT dynamic power spectrum. At around MJD 55000 till MJD 56000, the first torque reversal occurs, with the pulsar transitions from a spin-up to a spin-down phase. During this time the relative height of the fundamental frequency decreases to $\sim$ 5, and the peak corresponding to the second harmonic is absent, as also seen in Panel (a) of Figure \ref{fig:power_spectra}. Between MJD 58000 and MJD 59000, the fundamental period becomes weaker and consistent with the mean power. However, we see that the second harmonic, shown by a red dashed line, is present in the BAT power spectrum in Panel (b) of Figure \ref{fig:power_spectra} during MJD 58000 and MJD 59000, when the fundamental period is not strongly present. This is at the end of Spin Epoch II when the pulsar underwent a torque reversal from spin-up to spin-down behavior.
\par
The MAXI-GSC dynamic power spectrum, displayed in the middle of Figure \ref{fig:dynamic_power_spectra}b, shows the presence of the fundamental frequency of the superorbital modulations (black arrow). The fundamental frequency of the superorbital modulation is present, except for MJD 58000 till MJD 59000, similar to the trend observed in the BAT dynamic spectrum. However the second harmonic (depicted by the red arrow) is not strongly present around MJD 58000 till MJD 59000 as it is observed in the Swift--BAT spectrum. We also see several peaks in the MAXI-GSC dynamic power spectrum which are likely related to the precession period of the ISS. These peaks are not statistically significant and are below the 99.9\% white noise significance level. The non-detection of the second harmonic in the dynamic power spectrum constructed using MAXI-GSC data could be due to the low photon statistics. However we cannot rule out the possibility that the strength of the second harmonic is energy-dependent. The fundamental or the harmonic frequency of the superorbital modulations were not detected with a high statistical significance in the power spectra constructed from 2--4, 4--10 and 10--20 keV energy-resolved lightcurves of MAXI-GSC. Therefore we cannot further probe this energy dependence behaviour of the superorbital modulations. We can define three distinct time segments based on the behavior seen in the Swift-BAT dynamic power spectrum: Super Epoch A from MJD 53415 till MJD 57650, during which the fundamental frequency of the superorbital period was present and the second harmonic was not present, Super Epoch B from MJD 57650 till MJD 58800, during which the fundamental period had weakened and become consistent with the mean power and the second harmonic was present, and Super Epoch C from MJD 58800 till MJD 60750, during which the fundamental period reappeared and the second harmonic weakened and become consistent with the mean power. The significance at the peak of the superorbital period was estimated using the false-alarm probability (FAP; \citealt{scargle1982}) with values of $3 \times 10^{-8}$ for the Swift-BAT power spectrum and $5 \times 10^{-5}$ for the MAXI-GSC power spectrum for the entire lightcurves.
\par
\begin{figure}
    \centering
    \includegraphics[width=1\linewidth]{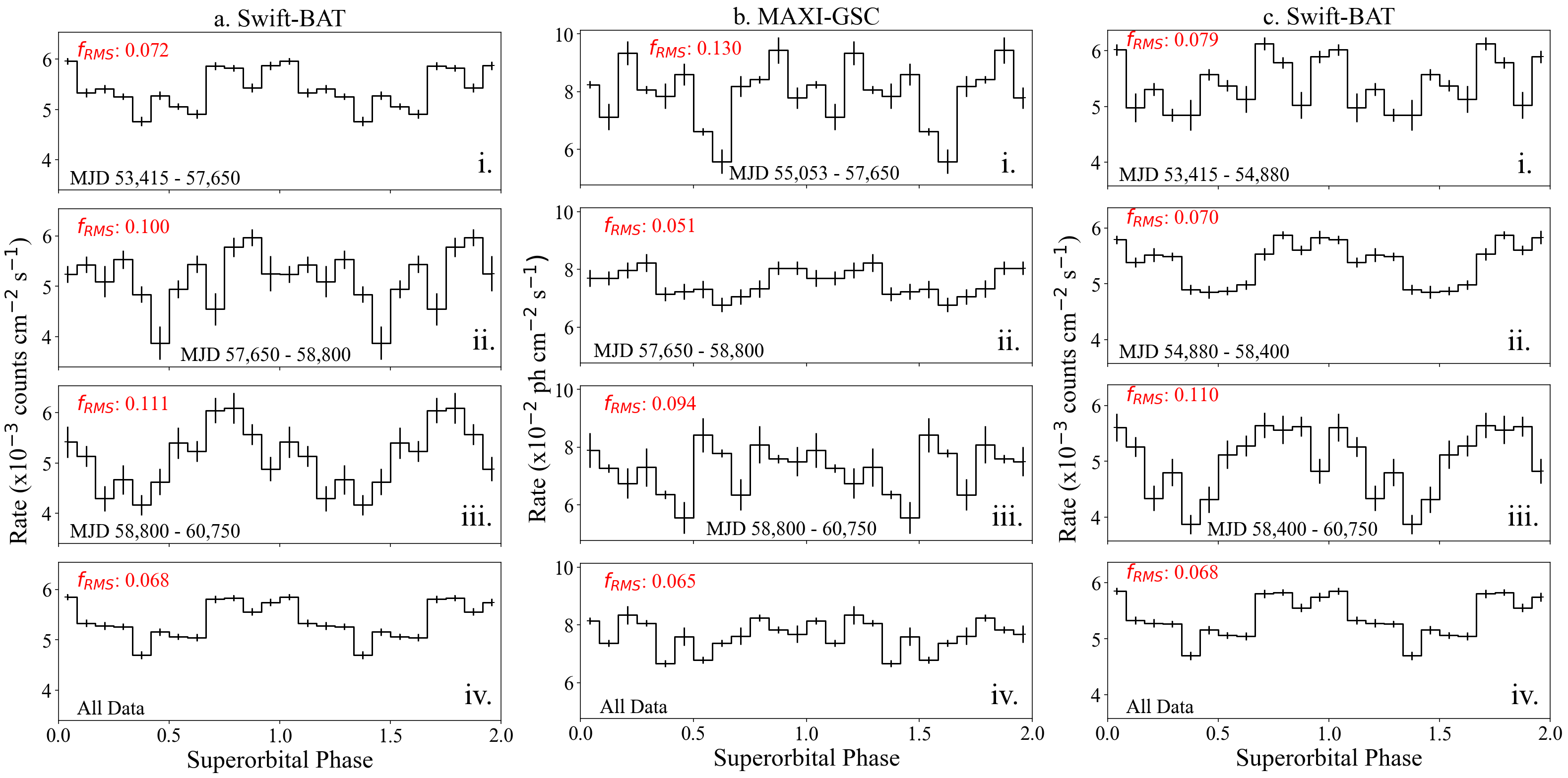}
    \caption{a: Different segments of the Swift--BAT lightcurves as defined in Section \ref{sec:dynamic power + intensity profiles}, folded on the superorbital period of 14.9081\,d. The fractional RMS ($f_{\rm{RMS}}$) was calculated for the superorbital intensity profiles. b: Segments of the MAXI-GSC lightcurves as defined in Section \ref{sec:dynamic power + intensity profiles}, folded in the same manner as Swift--BAT. c: Segments of Swift--BAT lightcurves, the data divided into epochs based on the torque reversals mentioned in Section \ref{sec:intro}. All lightcurves used MJD 56137.19, the epoch of maximum flux, as T$_0$.}
    \label{fig:swift and maxi epoch}
\end{figure}

To investigate the changes in the superorbital intensity profiles associated with a change in the strength of the superorbital modulation, we divided the Swift-BAT lightcurves into three different segments, those being Super Epoch A (MJD 53415 -- MJD 57650), Super Epoch B (MJD 57650 -- MJD 58800), and Super Epoch C (MJD 58800 -- MJD 60750). Figure \ref{fig:swift and maxi epoch}a shows the superorbital intensity profiles constructed from the Swift-BAT lightcurve during the respective Super Epochs, as well as the entire lightcurve (All Data). T$_{0}$ is chosen as MJD 56137.9 which is the time of the maximum flux for the superorbital intensity profile of All Data and this T$_{0}$ is used to fold all the Swift-BAT and MAXI-GSC lightcurves to create all superorbital intensity profiles. We divided the MAXI-GSC lightcurve into segments based on the same Super Epochs as we did Swift-BAT, although since MAXI-GSC observations began at MJD 55053, there is no data in Super Epoch A prior to that. Figure \ref{fig:swift and maxi epoch}b shows the superorbital intensity profiles constructed from the MAXI-GSC lightcurve during the aforementioned Super Epochs, as well as one constructed from the entire MAXI-GSC lightcurve (All Data). In addition, we also created superorbital intensity profiles which divided the Swift-BAT lightcurve into three segments defined by the Spin Epochs. Figure \ref{fig:swift and maxi epoch}c shows the Swift-BAT superorbital intensity profiles constructed from the Spin Epochs I, II, and III time intervals. We estimated the fractional root mean square (RMS) amplitude ($f_{\rm{RMS}}$) for all the superorbital intensity profiles using the formulation in \cite{Vaughan2003}.

\subsection{Pointed X-ray Observations}
\label{sec:Pointed X-ray Observations}

NuSTAR is a hard X-ray telescope operating in 3--79\,keV energy band \citep{Harrison2013}. It carries two co-aligned grazing-incidence Wolter I imaging telescopes that focus onto two independent focal plane modules, FPMA and FPMB. We carried out two NuSTAR observation around the predicted superobital maximum and minimum for the superorbital cycle, simultaneously with two NICER observations (PI: N.Islam). We also used two previous NuSTAR observations for analysis. \par
NICER is an X-ray observatory that has been operating on the International Space Station since 2017 \citep{Gendreau2016a}. The main instrument, the X-ray Timing Instrument (XTI), consists of 56 identical and coaligned cameras, each containing an X-ray Concentrator and an Si drift detector positioned in the concentrator's focal plane \citep{Gendreau2016b}. Along with the two NICER observations carried out simultaneously with NuSTAR observations, we used 8 previous NICER observations from MJD 59216 to MJD 59434. We also used results from spectral analysis of previous RXTE, INTEGRAL and Suzaku observations, to incorporate a larger range of superorbital and orbital period coverage.   
Table~\ref{Tab:observation_list} contains the details of the NuSTAR and NICER observations (arranged chronologically by the date of the observation and labeled as ``Instrument Obs Number"), and the archival Suzaku, RXTE and INTEGRAL observations (and the references to the literature where we use the results of spectral analysis) along with their superorbital and orbital phase.
\par
The NuSTAR observations were reduced and analyzed using NuSTAR Data Analysis Software (NuSTARDAS) version 2.1.2, {\tt HEASoft} version 6.34, and {\tt CALDB} version 20211020. The event files were reprocessed using {\tt nupipeline} using the standard filtering procedure and the default screening criteria. The lightcurve of the NuSTAR Obs I and NuSTAR Obs IV showed the presence of an X-ray eclipse and therefore we filtered out the X-ray eclipses from the event files, using X-ray eclipse filtered time with {\tt usrgti} as an input to {\tt nuproducts}. The event files were corrected to the solar system barycenter using {\tt barycorr} with the DE200 solar system ephemeris. The source spectra, response matrices, ancillary matrices and the energy resolved lightcurves (3--6\,keV, 6--10\,keV, 3--10\,keV, 10--20\,keV, 20--30\,keV, 30--70\,keV and 3--70\,keV) were extracted in the SCIENCE mode (01) from a circular region of radius 100'' centered on the source. The background lightcurves and spectra were extracted from a circular region of 100'' in a source-free region on the same chip. 
\par
We analyzed the NICER data using {\tt NICERDAS v13}, {\tt HEASoft v6.34}, and {\tt CALDB v20240206}. The standard pipeline processing was used to carry out screening and filtering of the data and apply the latest calibration files using {\tt nicerl2}, with an additional filtering criterion to remove non X-ray flares (NICER Obs VII, NICER Obs VIII and NICER Obs VI). The lightcurves of NICER Obs VII and NICER Obs VIII show the presence of an X-ray eclipse and we used a GTI file as an input in {\tt nicerl2} to filter out the X-ray eclipses to produce the cleaned event files. 
Using NICERDAS task {\tt nicerl3-lc} with the SCORPEON model as the background\footnote{\url{https://heasarc.gsfc.nasa.gov/docs/nicer/analysis_threads/nicerl3-lc/}}, we extracted background subtracted energy-resolved lightcurves (0.3--3\,keV, 3--6\,keV, 6--10\,keV, 3--10\,keV and 0.3--10\,keV), which were corrected to the solar system barycenter  with FTOOL {\tt barycorr}. We extracted the source, the appropriate response matrices, and ancillary files for NICER observations using NICERDAS task {\tt nicerl3-spect}. Both NuSTAR and NICER event files were corrected for the orbital motion of the NS in 4U 1538--52 using the ephemerides given in \cite{Hemphill2019}. The source spectra were rebinned using the \cite{KaastraBleeker2016} technique of optimal binning via the {\tt ftgrouppha} tool, with a minimum count number of 25 per bin. The spectral analysis for NICER spectra was carried out fitting the source spectrum and modeling the background spectrum using the SCORPEON background model\footnote{\url{https://heasarc.gsfc.nasa.gov/docs/nicer/analysis_threads/scorpeon-versions/}}. As NuSTAR Obs III (superorbital minimum) and NuSTAR Obs IV (superorbital maximum) were conducted simultaneously with NICER Obs IV and NICER Obs VI, respectively, we carry out joint broadband X-ray spectral fitting of these observations (discussed in detail in Section~\ref{sec:spectral analysis}).
\par
The NICER and NuSTAR observations of 4U 1538--52 (as well as archival RXTE, Suzaku, and INTEGRAL observations) are distributed mostly evenly over the orbital period with a wide phase coverage, with none carried out between orbital phases 0.3 to 0.4. The pointed X-ray observations were carried between superorbital phases 0 to 1.0, except for no observations in phases 0.2-0.5, and a majority of the observations between phase 0.9 and 1.0. Two INTEGRAL observations were carried out in Spin Epoch I of pulse period changes, one INTEGRAL observation, the Suzaku, and two NuSTAR observations were carried out during Spin Epoch II, and two NuSTAR and all NICER observations were carried out during Spin Epoch III.

\subsubsection{Pulsation analysis}

\begin{table}[!ht]
    \centering
    \caption{NuSTAR and NICER
observations of 4U 1538--52 arranged chronologically by the date of the observation and labeled as “Instrument Obs Number". X-ray observations whose results are used are also included along with their references.}
    \rotatebox{90}{
    \begin{threeparttable}
    \begin{tabular}{lllllllll}
    \hline
        OBSID & Label & Time Start (MJD) & Exposure (ks) & Pulse Period (s) & Superorbital Phase & Orbital phase & ~ \\ \hline
        30201028002 & NuSTAR Obs I & 57611.799 & 10.4 & 526.94 \textpm 0.14 & 0.945 - 0.013 & 0.804 - 0.927 \\
        30401025002 & NuSTAR Obs II & 58605.938 & 36.9 & 527.00 \textpm 0.03 & 0.608 - 0.668 & 0.450 - 0.691 \\
        30602024002 & NuSTAR Obs III & 59261.327 & 21.8 & 526.00 \textpm 0.27 & 0.556 - 0.591 & 0.237 - 0.379 \\
        30602024004 & NuSTAR Obs IV & 59267.029 & 13.7 & 526.61 \textpm 0.29 & 0.938 - 0.975 & 0.766 - 0.914 \\ \hline
        3667010101 & NICER Obs I & 59216.047 & 7.0 & 526.10 \textpm 0.31 & 0.519 - 0.530 & 0.092 - 0.134 \\
        3667010201 & NICER Obs II & 59221.916 & 2.4 & ~ & 0.913 - 0.915 & 0.666 - 0.675 \\
        3667010202 & NICER Obs III & 59221.981 & 4.8 & 526.93 \textpm 0.34 & 0.917 - 0.923 & 0.684 - 0.709 \\
        3667010301 & NICER Obs IV & 59261.367 & 7.3 & 525.61 \textpm 0.32 & 0.558 - 0.589 & 0.248 - 0.373 \\
        3667010302 & NICER Obs V & 59267.032 & 0.3 & ~ & 0.938 - 0.943 & 0.767 - 0.789 \\
        3667010401 & NICER Obs VI & 59267.163 & 1.0 & ~ & 0.947 - 0.974 & 0.802 - 0.910 \\
        4594010101 & NICER Obs VII & 59431.097 & 1.8 & ~ & 0.940 - 1.000 & 0.772 - 0.014 \\
        4594010102 & NICER Obs VIII & 59432.009 & 10.3 & 526.00 \textpm 0.30 & 0.001 - 0.066 & 0.017 - 0.278 \\
        4594010103 & NICER Obs IX & 59433.01  & 14.9 & 526.21 \textpm 0.29 & 0.068 - 0.134 & 0.288 - 0.551 \\
        4594010104 & NICER Obs X & 59434.051 & 15.1 & 527.00 \textpm 0.30 & 0.144 - 0.195 & 0.565 - 0.794 \\ \hline

        ID & X-ray mission & ~ & Exposure (PCA) & ~ & ~ & ~ & Reference \\ \hline
        10145 & RXTE & 50450.620 & 114.1 & 528.824 \textpm 0.014 & 0.748 - 0.950 & 0.047 - 0.854 & \cite{Hemphill2016} \\
        20146 & RXTE & 50411.960 & 56.4 & 527.978 \textpm 0.001 & 0.156 - 0.851 & 0.677 - 0.456 & \textquotedbl \\
        50067 & RXTE-PCA & 51924.880 & 99.1 & 527.60 \textpm 0.01 & 0.606 - 0.841 & 0.470 - 0.411 & \textquotedbl \\
        80016 & RXTE & 52851.950 & 53.4 & 526.83 \textpm 0.01 & 0.771 - 0.200 & 0.127 - 0.843 & \textquotedbl \\ \hline
        ~ & ~ & ~ & Exposure (JEM-X 1) & ~ & ~ & ~ \\ \hline
        0200–0299 & INTEGRAL & 53198.100 & 84.6 & 526.401 \textpm 0.002 & 0.982 - 0.163 & 0.970 - 0.691 & \textquotedbl \tnote{a}\\
        0300–0399 & INTEGRAL & 53465.100 & 39.6 & 526.644 \textpm 0.003 & 0.886 - 0.333 & 0.584 - 0.373 & \textquotedbl \\
        0700–0799 & INTEGRAL & 54747.900 & 77.2 & 525.262 \textpm 0.002 & 0.905 - 0.022 & 0.655 - 0.122 & \textquotedbl \\
        0900–0999 & INTEGRAL & 55252.700 & 20.5 & 525.401 \textpm 0.004 & 0.755 - 0.182 & 0.051 - 0.761 & \textquotedbl \\ \hline
        ~ & ~ & ~ & Exposure (XIS) & ~ & ~ & ~ \\ \hline
        407068010 & Suzaku & 56149.020 & 46.0 & 525.59 \textpm 0.04 & 0.858 - 0.906 & 0.460 - 0.651 & \textquotedbl \\ \hline

    \end{tabular}
    \begin{tablenotes}
\item[a] \qquad \qquad \qquad a) Pulse periods of INTEGRAL observations are taken from \cite{Hemphill2013}.
\end{tablenotes}
\end{threeparttable}
}
    \label{Tab:observation_list}
\end{table}

\begin{figure}
    \centering
    \includegraphics[width=0.7\linewidth]{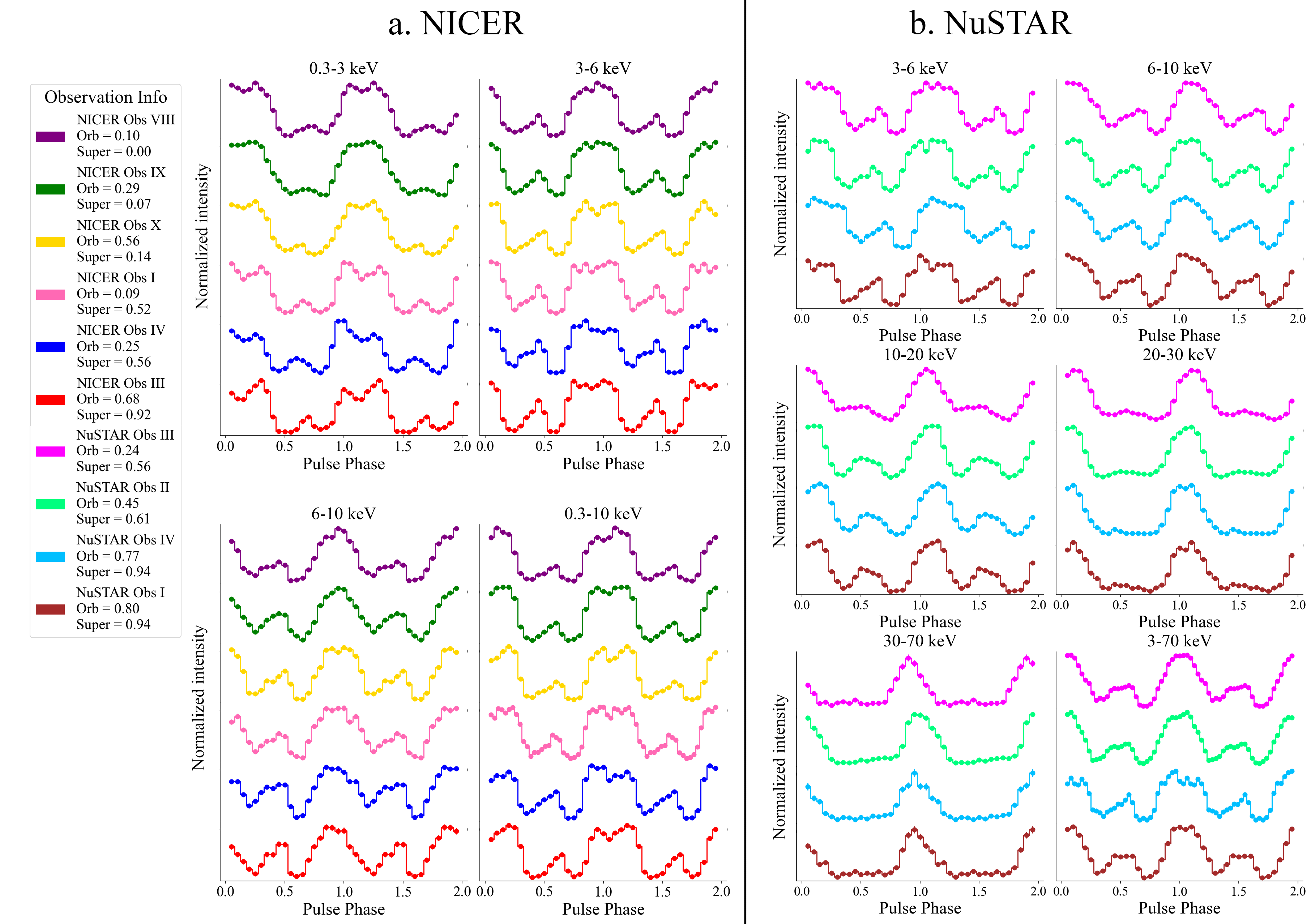}
    \caption{Left: The pulse profiles of the NICER observations of 4U 1538--52 arrayed to show the changes in pulse profiles as a function of orbital and superorbital phase. The folded pulse profiles are normalized by dividing by the average source intensity to create Normalized Intensity plotted on the y-axis. 
    The observations are arrayed in order of increasing superorbital phase. Right: The pulse profiles of the NuSTAR observations of 4U 1538--52 in the different energy bands. The individual pulse profiles are given an arbitrary shift in the y-axis to enhance clarity.}
    \label{fig:NICER NuSTAR offset}
\end{figure}

\begin{figure}
    \centering
    \includegraphics[width=0.95\linewidth]{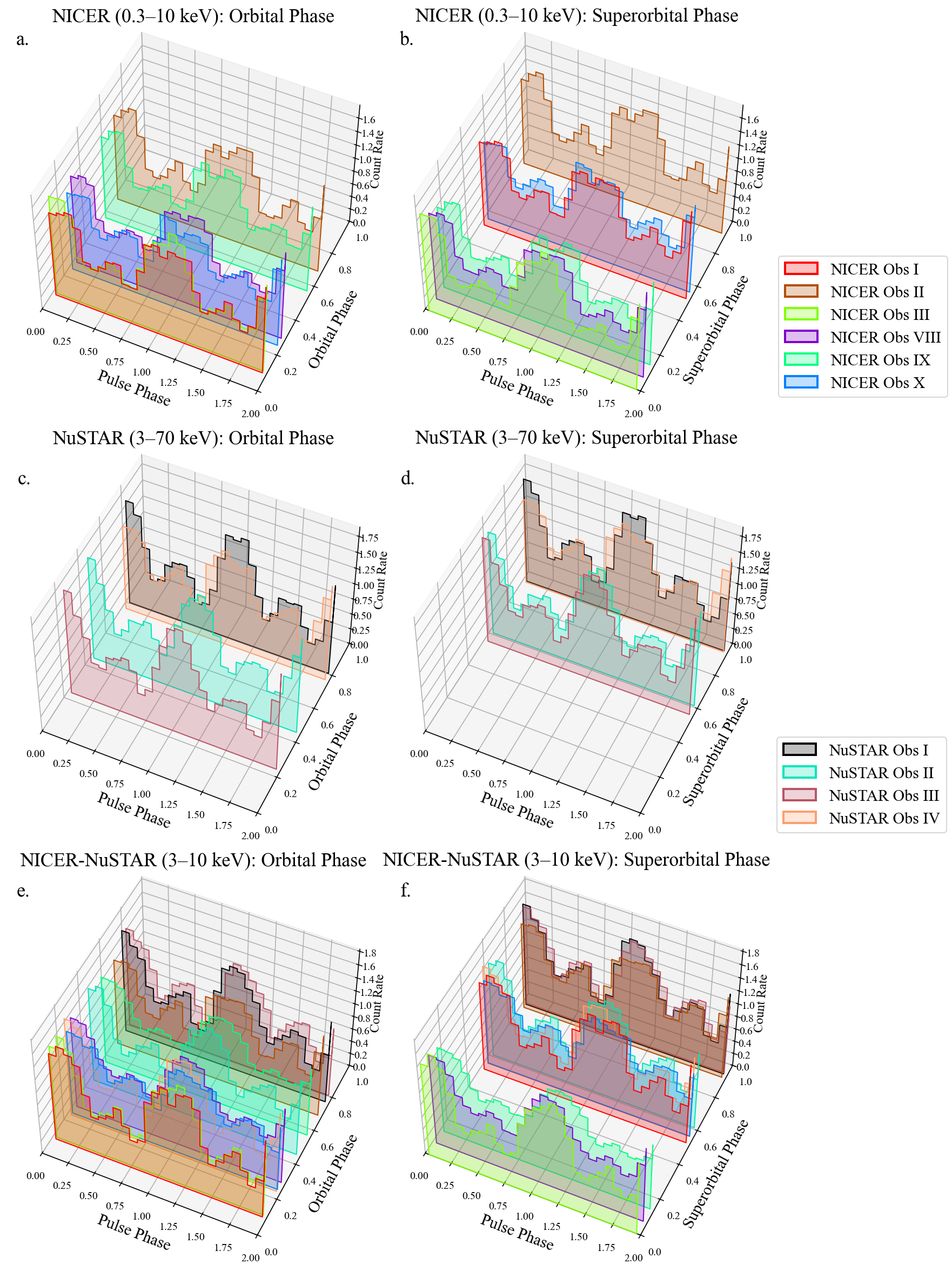}
    \caption{Pulse profiles of 4U 1538--52 constructed from the NICER observations in 0.3--10\,keV (a and b) and NuSTAR observations (c and d) in 3--70\,keV energy bands, showing the count-rate (in units of counts $\text{cm}^{-2}$ $\text{s}^{-1}$), pulse phase, and orbital/superorbital phase. e and f: Pulse profiles in the common 3-10 keV energy band from NICER and NuSTAR observations.}
    \label{fig:3D NICER NuSTAR}
\end{figure}

We searched for pulsations in the NICER lightcurves in 0.3--10\,keV and NuSTAR lightcurves in 3--70\,keV energy band using epoch folding and $\chi^{2}$ maximization techniques using the FTOOL {\tt efsearch} \citep{Leahy1983, Leahy1987}. Only NICER observations which have an effective exposure greater than 4.4\,ks were used for this analysis; observations that did not meet this exposure threshold did not possess enough counts for meaningful pulsation studies due to the length of the NS's rotation period. The pulse periods and their respective error-bars for the individual observations are listed in Table \ref{Tab:observation_list}. The pulse periods were found using FTOOL {\tt efsearch}, and errors were determined using the bootstrap method, where simulated lightcurves were generated in which the count rate at each instant of time was determined from $y_{i}' = y_{i} +\sigma_{y{i}}$, where $y_{i}'$ is the count rate at a particular instant in the simulated lightcurve, $y_{i}$ is the count rate at that instant in the original lightcurve, and $\sigma_{y{i}}$ is a quantity with a Gaussian distribution in the interval between ($-3\sigma,3\sigma$) (with $\sigma$ being the standard deviation of the original lightcurve). We then analyzed the simulated lightcurve and obtained the pulse period using the same methods as we did the original; repeating this process for 1,000 simulated lightcurves creates a vector of pulse periods for different realizations of the original lightcurve, the standard deviation of which is the estimate of the period's error \citep{Boldin2013}.

We used the FTOOL {\tt efold} to fold the energy resolved light-curves over the estimated pulse periods for each observation for 20 phasebins and obtained the pulse profiles for the following energy ranges: 0.3--3\,keV, 3--6\,keV, 6--10\,keV, 3--10\,keV, and 0.3--10\,keV for NICER, and 3--6\,keV, 6--10\,keV, 10--20\,keV, 20--30\,keV, 30--70\,keV, and 3--70\,keV for NuSTAR. The pulse profile obtained from a given observation within a specific energy range is compared with those from other observations in the same energy range by aligning their shapes using a CrossCorrelation function\footnote{\url{https://docs.stingray.science/en/v1.1.2.4/notebooks/CrossCorrelation/cross_correlation_notebook.html}} in {\tt Stingray v1.1.2}. Energy-resolved pulse profiles for each observation are presented in Figure \ref{fig:NICER NuSTAR offset}. Figure \ref{fig:3D NICER NuSTAR} is a set of three dimensional plots of the NICER and NuSTAR pulse profiles in the 0.3--10\,keV energy band (NICER), 3--70\,keV energy band (NuSTAR), and 3-10\,keV (both NICER and NuSTAR). These depict the pulse phase on the x-axis, the orbital (panels a, c, and e) or superorbital (panels b, d, and f) phase on the y-axis, and the count rate on the z-axis. Using Figure \ref{fig:3D NICER NuSTAR}, we examine variations in the shape of the pulse profiles as a function of their orbital and superorbital phases. 

\subsection{Spectral Analysis}
\label{sec:spectral analysis}

\begin{sidewaystable}
\caption{Best-fitting spectral parameter values using the models defined in Section \ref{sec:Res_Spectral Analysis} for the NICER and NuSTAR observations. Errors are calculated at the 90\% confidence limit.}
\label{tab:NICER and NuSTAR spectral fit values}
\hspace*{-7cm}
\begin{minipage}{\textheight}
\begin{threeparttable}
\begin{tabular}{c | m{1.6cm} | m{1.6cm} | m{1.4cm} | m{1.5cm} |
m{1.4cm} | m{1.5cm} | m{1.6cm} | m{1.4cm} | m{1.5cm} |
m{1.7cm} | m{1.7cm} | c | c}
\hline
Observation &
\shortstack{C\textsubscript{XTI}} &
\shortstack{C\textsubscript{FPMB}} &
\shortstack{N\textsubscript{H\textsubscript{1}}\tnote{*}} &
\shortstack{N\textsubscript{H\textsubscript{2}}\tnote{*}} &
\shortstack{pcf} &
\shortstack{$\Gamma$} &
\shortstack{E\textsubscript{C}\\(keV)} &
\shortstack{E\textsubscript{F}\\(keV)} &
\shortstack{E\textsubscript{cyc}\\(keV)} &
\shortstack{F\textsubscript{NICER}$^{\dagger}$\tnote{a}} &
\shortstack{F\textsubscript{NuSTAR}$^{\dagger}$\tnote{b}} &
$\chi^{2}$ &
d.o.f.\tnote{c} \\
\hline
NICER Obs I & ~ & ~ & $2.9_{-0.3}^{+0.2}$ & $6.0_{-0.9}^{+1.1}$ & $0.67 \pm 0.05$ & $1.2 \pm 0.2$ & ~ & ~ & ~ & $3.65 \pm 0.03$ & ~ & 75.03 & 111 \\
NICER Obs II & ~ & ~ & $1.48 \smash{\pm} 0.06$ & $10_{-2}^{+3}$ & $0.1 \pm 0.1$ & $1.3_{-0.1}^{+0.2}$ & ~ & ~ & ~ & $4.32 \pm 0.03$ & ~ & 106.15 & 104 \\
NICER Obs III & ~ & ~ & $1.2_{-0.3}^{+0.2}$ & $1.1_{-0.3}^{+1.1}$ & $0.6 \pm 0.3$ & $1.01_{-0.09}^{+0.08}$ & ~ & ~ & ~ & $3.21 \pm 0.03$ & ~ & 133.02 & 101 \\
NICER Obs VI & ~ & ~ & $0.9_{-0.3}^{+0.4}$ & $8.4 \pm 0.4$ & $0.88_{-0.05}^{+0.03}$ & $0.5 \pm 0.2$ & ~ & ~ & ~ & $7.2_{-0.6}^{+0.8}$ & ~ & 98.54 & 101 \\
NICER Obs VII & ~ & ~ & $1.4_{-0.2}^{+0.4}$ & $3.8_{-0.3}^{+0.5}$ & $0.86_{-0.06}^{+0.03}$ & $1.0_{-0.1}^{+0.2}$ & ~ & ~ & ~ & $3.94_{-0.04}^{+0.05}$ & ~ & 127.32 & 115 \\
NICER Obs VIII & ~ & ~ & $1.61_{-0.07}^{+0.04}$ & $5.7_{-0.6}^{+0.5}$ & $0.52_{-0.04}^{+0.02}$ & $1.3 \pm 0.1$ & ~ & ~ & ~ & $4.71 \pm 0.02$ & ~ & 95.46 & 126 \\
NICER Obs IX & ~ & ~ & $1.40_{-0.03}^{+0.04}$ & $10_{-2}^{+3}$ & $0.19 \smash{\pm} 0.06$ & $1.35_{-0.07}^{+0.08}$ & ~ & ~ & ~ & $4.69 \pm 0.02$ & ~ & 104.70 & 132 \\
NICER Obs X & ~ & ~ & $1.61 \smash{\pm} 0.08$ & $3 \pm 1$ & $0.31_{-0.04}^{+0.03}$ & $1.15_{-0.05}^{+0.10}$ & ~ & ~ & ~ & $4.24 \pm 0.02$ & ~ & 121.33 & 131 \\

NuSTAR Obs I & ~ & $0.97 \pm 0.01$ & $8.9_{-0.6}^{+0.3}$ & ~ & ~ & $1.06 \pm 0.03$ & $14.5_{-0.4}^{+0.5}$ & $12.3_{-0.5}^{+0.6}$ & $22.0 \pm 0.3$ & ~ & $6.95 \pm 0.07$ & 288.52 & 262 \\
NuSTAR Obs II & ~ & $0.98 \pm 0.01$ & $0.8 \pm 0.2$ & ~ & ~ & $1.19 \pm 0.01$ & $14.7 \pm 0.1$ & $11.8 \pm 0.2$ & $21.8 \pm 0.1$ & ~ & $9.46 \pm 0.02$ & 502.95 & 339 \\
NuSTAR Obs III & $1.02 \pm 0.02$ & $1.01 \pm 0.01$ & $1.4_{-0.7}^{+0.1}$ & $1.0 \pm 0.5$ & $0.24 \pm 0.1$ & $1.11 \pm 0.02$ & $15.8_{-0.4}^{+0.5}$ & $10.6_{-0.5}^{+0.6}$ & $21.6 \pm 0.2$ & $3.61 \pm 0.02$ & $5.45 \pm 0.02$ & 408.03 & 402 \\

NuSTAR Obs IV & $1.04 \pm 0.07$ & $1.01 \pm 0.01$ & $0.8_{-0.4}^{+0.6}$ & $9.1_{-0.7}^{+1.5}$ & $0.93_{-0.03}^{+0.02}$ & $0.97_{-0.06}^{+0.07}$ & $13.5 \pm 0.3$ & $11.2_{-0.4}^{+0.5}$ & $21.9 \pm 0.2$ & $3.5_{-0.2}^{+0.1}$ & $5.96_{-0.05}^{+0.03}$ & 388.15 & 331 \\

\hline
\end{tabular}
\begin{tablenotes}
\item[*] \qquad \qquad \qquad \qquad \qquad \qquad \qquad *) in units of $\times 10^{22}$ cm$^{-2}$.
\item[$\dagger$] \qquad \qquad \qquad \qquad \qquad \qquad \qquad $\dagger$) in units of $\times 10^{-10}$ erg s$^{-1}$ cm$^{-2}$.
\item[a] \qquad \qquad \qquad \qquad \qquad \qquad \qquad a) unabsorbed flux for NICER uses energy range 0.3--10.0\,keV.
\item[b] \qquad \qquad \qquad \qquad \qquad \qquad \qquad b) unabsorbed flux for NuSTAR uses energy range 3.0--70.0\,keV.
\item[c] \qquad \qquad \qquad \qquad \qquad \qquad \qquad c) degrees of freedom.
\end{tablenotes}

\end{threeparttable}
\end{minipage}
\end{sidewaystable}

\begin{figure}
    \centering
    \includegraphics[width=\linewidth]{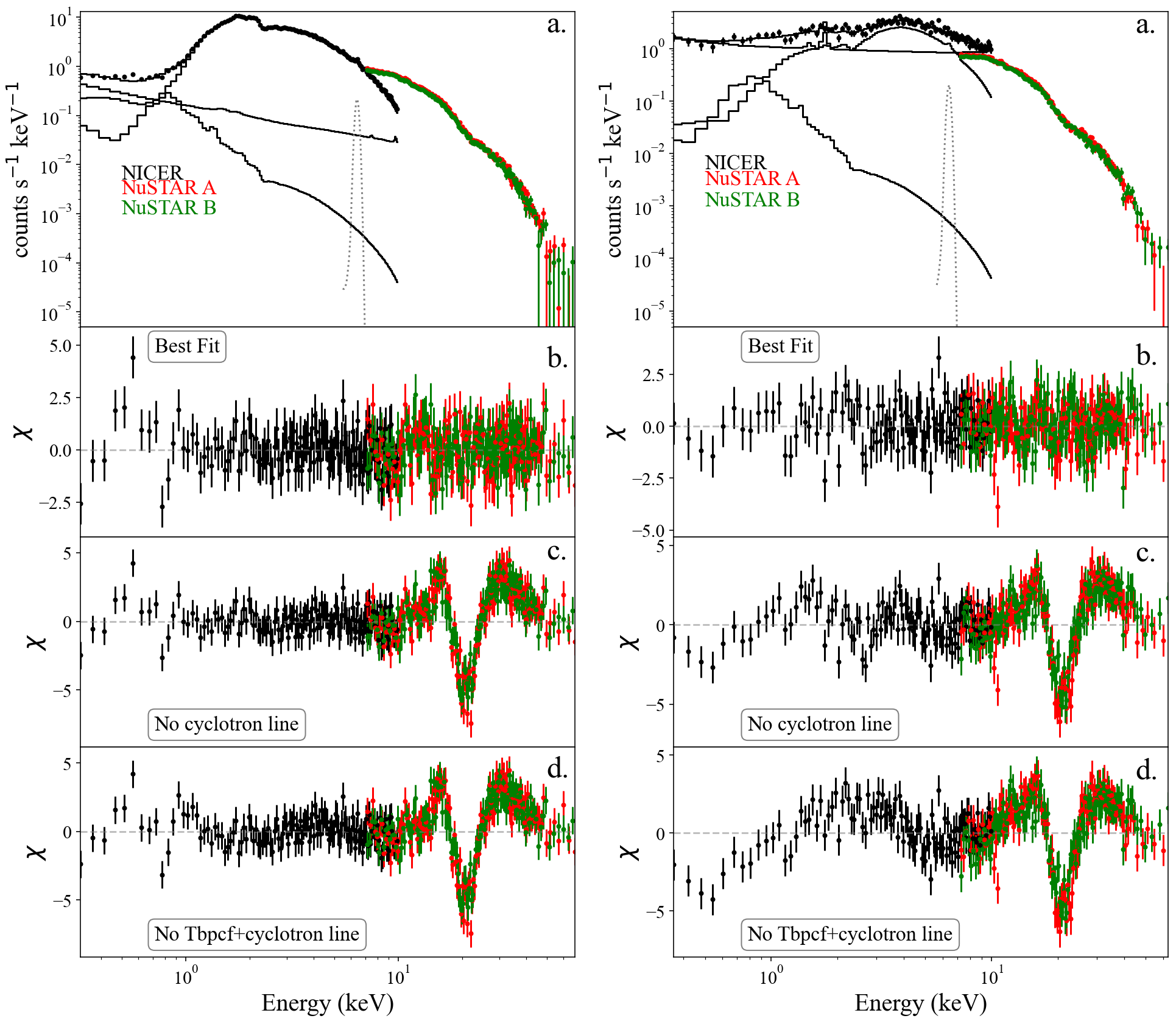}
    \caption{Left a: The spectral fit of NICER Obs IV and NuSTAR Obs III. The NICER observations are in energy band 0.3--10\,keV, and the NuSTAR observations are in 7--70\,keV. Left b: The residuals of the spectral fit of NICER Obs IV and NuSTAR Obs III. Left c: The residuals of the spectral fit of NICER Obs IV and NuSTAR Obs III when the cyclotron line is removed. Left d: The residuals of the spectral fit of NICER Obs IV and NuSTAR Obs III when both the cyclotron line and the partial covering absorption models are removed. Right a: The spectral fit of NICER Obs V and NuSTAR Obs IV. Right b: The residuals of the spectral fit of NICER Obs V and NuSTAR Obs IV. Right c: The residuals of the spectral fit of NICER Obs V and NuSTAR Obs IV when the cyclotron line is removed. Right d: The residuals of the spectral fit of NICER Obs V and NuSTAR Obs IV when both the cyclotron line and partial covering absorption models are removed.}
    \label{fig:NICER NuSTAR spectral fits}
\end{figure}

The X-ray broadband fitting of the simultaneous NICER and NuSTAR spectra (NICER Obs IV with NuSTAR Obs III and NICER Obs V with NuSTAR Obs IV) was carried out in 0.3--10\,keV for NICER and 7--70\,keV for NuSTAR, since there are uncertainties in the cross-calibration between NICER and NuSTAR up to 7\,keV \footnote{\url{https://heasarc.gsfc.nasa.gov/docs/nicer/data_analysis/nicer_analysis_tips.html}} and above 70 keV the background count rate dominates the source count rate for the NuSTAR spectra. All spectral fits were carried out using {\tt XSPEC v12.15.0d}. For modeling the X-ray continuum spectra, we employ a modified power-law with a cutoff model {\tt mplcut} consisting of a power-law with a high energy cutoff and a narrow Gaussian absorption line with its energy tied to the cutoff energy to ``smooth" the introduction of the cutoff, as used by \cite{Coburn2001, Hemphill2019}. The continuum model was modified by a photoelectric absorption component that fully covers the source {\tt Tbabs} and a partial covering absorption {\tt Tbpcf}. The fundamental and harmonic of the cyclotron line is modeled by two Gaussian absorption lines {\tt gabs}, where the line energy of the harmonic is fixed to 50\,keV \citep{Rodes-Roca2009} since it was poorly constrained in the NuSTAR spectra. The Fe K$\alpha$ and K$\beta$ fluorescence lines were modeled by two Gaussian emission lines {\tt gaus}. For the spectral fits where the centroid energy of the gaussians were inaccurate, the line energies were fixed to 6.4 and 7.1\,keV respectively. The line widths of the gaussians were fixed to 0.1\,keV for all observations, since keeping them free led to unphysical values. The cross-calibration constants normalized to NuSTAR-FPMA were multiplied with the above spectral models to account for the instrumental calibration uncertainties. The spectral model is defined below:\\
{\tt constant * Tbabs * Tbpcf * (mplcut * gabs(fundamental) * gabs(harmonic) + gaus (6.4\,keV) + gaus (7.1\,keV))} \\
 Figure \ref{fig:NICER NuSTAR spectral fits} show the spectral fits to the joint NICER and NuSTAR spectra for the best fit spectral model. The bottom panels show the residuals of the best fit model, the best fit model when the {\tt gabs} which models the cyclotron line is excluded, and the best fit model when excluding both the cyclotron line and partial covering absorber models.  
\par
We fit NuSTAR Obs I and II with the above model, excluding the partial covering absorber {\tt Tbpcf}. The NICER observations (I, II, III, VI, VII, VIII, IX and X) were fitted with the spectral model consisting of a power-law, modified by a photoelectric absorption component that fully covers the source {\tt Tbabs} and two Gaussian emission lines to model the Fe K$\alpha$ and K$\beta$ fluorescence lines, with the energies and line widths fixed to the values used for the joint NICER and NuSTAR spectral fits. The results of the spectral fits are reported with errors at the 90\% confidence limit. The unabsorbed X-ray fluxes and their errors were estimated using the {\tt cflux} model in XSPEC. The results are reported in Table \ref{tab:NICER and NuSTAR spectral fit values}. We plot the spectral parameters as functions of the orbital and superorbital phases in Figure \ref{fig:prev obs orbital + superorbital wINTEGRAL}. We also plot the spectral parameters of RXTE, Suzaku, and INTEGRAL spectral fittings which used the same model as above.

\section{Results and Discussions}
\label{sec:Results}
\subsection{Results from Swift-BAT, MAXI-GSC and Fermi-GBM Observations}
\label{sec:Res_Long-Term Observations}

The top panel of Figure \ref{fig:4u1538 fermi} shows the pulse-period history of 4U 1538--52 from Fermi-GBM, RXTE, and INTEGRAL data. The middle and bottom panels of Figure \ref{fig:4u1538 fermi} are the Swift-BAT and MAXI-GSC lightcurves rebinned with the superorbital period, which suggests that the pulse period changes are not related to the changes in the X-ray flux of the source. We also confirm no change in the GBM pulsed flux was observed with changes with the pulse period. No significant long-term intensity variations are found between different superorbital cycles, consistent with that seen for other sgHMXBs showing superorbital modulations \citep{Coley2019, Islam2023}. 
\par
Figure \ref{fig:dynamic_power_spectra} shows the dynamic power spectra constructed from Swift-BAT and MAXI-GSC lightcurves in the 15--50\,keV and 2--20\,keV energy bands respectively. The power spectra constructed using both all-sky monitors show a prominent peak at the fundamental frequency of superorbital period at relative power of $\geq$ 20 (marked by a black arrow) and which is larger than the 99.9\% and 99.99\% of significance level of the white noise. This confirms that the fundamental frequency of the superorbital modulations are present in the MAXI-GSC lightcurves at the 2--20\,keV energy-band, in addition to the 15--50\,keV Swift-BAT lightcurves. The red arrow marks the second harmonic of the superorbital modulations. The peak corresponding to the second harmonic is not strongly present in the average power spectrum. 
\par
The middle panels show the dynamic power spectra constructed using 750\,d time spans with a sliding window offset of 50\,d, as mentioned in Section \ref{sec:dynamic power + intensity profiles}. Since consecutive segments in the dynamic power spectra overlap, the resulting measurements are not statistically independent, and therefore the precise epochs of modulation changes cannot be unambiguously determined. The Swift-BAT dynamic power spectrum reveals the strength of the fundamental frequency from MJD 53415 to 57650 with a relative height of $\sim$ 10. However, around MJD 55000 till MJD 56000, during which the first torque reversal — when the pulsar transitions from a spin-up to a spin-down phase — takes place, the relative height of the fundamental frequency decreases to $\sim$ 5. The peak corresponding to the second harmonic is absent in this time period. As noted in \cite{corbet2021}, the peak at the fundamental frequency weakens and becomes consistent with the mean power around MJD 57650 till MJD 58800, while the second harmonic becomes stronger during the same interval. During this period, we find the second harmonic's relative height to be about $\sim$2 in Swift-BAT dynamic power spectra. A similar behavior, where either or both the fundamental and second harmonic components are observed, has been reported in other sgHMXBs exhibiting superorbital modulations, such as 4U 1909+07, IGR J16393--4643, and IGR J16479--4514 \citep{Islam2023}. The weakening of the peak at the fundamental and the presence of a stronger second harmonic accompanies the second torque reversal, marking the transition of the pulsar from a spin-down to a spin-up phase. We however note that exact temporal relation between the epoch of the torque reversal and changes in the strength of fundamental and harmonic frequencies cannot be ascertained due to the 750\,d length, and the 50\,d steps of the sliding window used in constructing the dynamic power spectrum. Similar variations in the amplitude of superorbital modulation in relation to its spin behavior have been observed in 2S 0114+650, another sgHMXB exhibiting superorbital modulations \citep{Hu2017}. Continued long-term monitoring of these systems is essential to establish a definitive link between superorbital modulation and pulse period changes.
\par
The MAXI-GSC dynamic power spectrum (middle panel of Figure \ref{fig:dynamic_power_spectra}b) displays the strength of the fundamental period of the superorbital modulation; however, the signal is relatively noisy and shows a lower relative height of $\sim$5, compared to that observed in the Swift-BAT dynamic power spectrum. Given the noisy nature of the power spectrum, we refrain from drawing definitive conclusions regarding variations in the superorbital modulation based on the MAXI-GSC power spectrum, and instead rely on the Swift-BAT dynamic power spectrum to trace changes in the strength of the superorbital modulation. Nevertheless, the detection of superorbital modulation in the MAXI-GSC lightcurves confirms that the modulations are also present in the lower X-ray energy range of 2--20\,keV. Superorbital modulation has generally not been detected in MAXI-GSC observations of other sgHMXBs that display such behavior in Swift-BAT lightcurves; it is likely that was first detected in 4U 1538--52 because it is the brightest source among this sample.
\par
Figure \ref{fig:swift and maxi epoch}a and b show the superorbital intensity profiles using Swift-BAT and MAXI-GSC lightcurves in the different time segments when either the fundamental or the second harmonic of the superorbital modulations are present, which were defined in Section \ref{sec:dynamic power + intensity profiles} as Super Epochs A, B, and C. The Swift--BAT superorbital intensity profiles for all the segments show similar profiles with a broad maximum around phase zero. There is a shift in the phase of the maximum flux for the intensity profile of Super Epoch B, when the second harmonic is present and the fundamental frequency reappears (Figure \ref{fig:swift and maxi epoch}a-ii). The MAXI--GSC superorbital intensity profiles do not show as pronounced a shape as the BAT profiles, probably due to the fact that the superorbital modulations are weaker in MAXI-GSC than in Swift--BAT. Figure \ref{fig:swift and maxi epoch}c shows the Swift--BAT superorbital intensity profiles in the three epochs of pulse period behavior changes. Spin Epoch I (Figure \ref{fig:swift and maxi epoch}c-i) corresponds to a short time interval and the Swift--BAT profile does not exhibit any distinct or well-defined structure. We notice a change in the shape of the profiles and the $f_{\rm{RMS}}$ between Spin Epoch II (Figure \ref{fig:swift and maxi epoch}c-ii) and Spin Epoch III (Figure \ref{fig:swift and maxi epoch}c-iii), suggesting a dependence on the superorbital modulations and pulse period changes in 4U 1538--52.  

\subsection{Pulsation Studies}
\label{sec:Res_Pulsation Studies}

To investigate the changes in the pulsation characteristics as a function of orbital and superorbital phases, we create energy resolved pulse profiles from the NICER and NuSTAR observations. The NICER pulse profiles are constructed in the 0.3--3\,keV, 3--6\,keV, 6--10\,keV and 0.3--10\,keV energy bands and are shown in Figure \ref{fig:NICER NuSTAR offset}a. These NICER energy-resolved pulse profiles show a similar double-peaked profile at different orbital and superorbital phases: a primary peak at pulse phase zero and a secondary smaller peak at pulse phase 0.5. The NuSTAR pulse profiles are constructed in the 3--6\,keV, 6--10\,keV, 10--20\,keV, 20--30\,keV, 30--70\,keV and 3--70\,keV energy bands, and are shown in Figure \ref{fig:NICER NuSTAR offset}b. These NuSTAR energy resolved pulse profiles show a change in the shape of the pulse profiles as a function of energy: it changes from a double-peaked profile to a single peak at energies above 30\,keV. The 0.3--10\,keV energy band NICER pulse profiles, the 3--70\,keV energy band NuSTAR pulse profiles, and the 3--10\,keV energy band NICER and NuSTAR pulse profiles are shown in more detail in Figure \ref{fig:3D NICER NuSTAR}, displaying the distribution of the observations over the orbital and superorbital periods. To compare pulse profiles for a wider range of orbital and superorbital phases, we create energy resolved pulse profiles in the common energy band of 3--10\,keV for NuSTAR and NICER observations. As displayed in Figure \ref{fig:3D NICER NuSTAR}e and f, we do not see any noticeable change in the pulse profile as a function of superorbital or orbital phase, which suggests the mechanisms driving the superorbital modulations are not related to the emission processes at the poles of the NS. We also confirm there are no changes in the pulse profiles in comparison to the pulse profiles reported in \cite{Hemphill2013, Hemphill2016} for previous RXTE and Suzaku observations, \cite{Varun2019} for AstroSAT observations, and \citep{Robba2001} for BeppoSAX observations. \cite{Hu2024} found no changes in the shape of the pulse profiles above 3\,keV for NuSTAR observations carried out in Spin Epoch II and Spin Epoch III. We extend the investigations using the NICER observations and find no changes in the shape of pulse profiles from 0.3\,keV to 70\,keV compared to its pulse period changes and no changes in the X-ray flux, suggesting the pulse period changes observed in 4U 1538--52 are not related to the presence of a prograde- or retrograde- accretion disk in the system \citep{Nelson1997}.

\subsection{Spectral Analysis}
\label{sec:Res_Spectral Analysis}

For accreting X-ray pulsars that exhibit cyclotron lines in their X-ray spectra, accurate modeling of the X-ray continuum is essential for reliable determination of the cyclotron line parameters \citep{Muller2013}. The simultaneous NICER and NuSTAR observations provide an opportunity to perform broadband X-ray spectral modeling, enabling robust constraints on both the soft and hard X-ray continuum. We carried out joint spectral fits using a model composed of a modified power-law with high-energy cutoff ({\tt mplcut}) and two photoelectric absorption components: a fully covering {\tt Tbabs} and a partially covering {\tt Tbpcf}. The fundamental and second harmonic cyclotron lines were modeled using two Gaussian absorption components ({\tt gabs}), while the Fe K$\alpha$ and K$\beta$ fluorescence lines were represented by Gaussian emission {\tt gaus} components. This choice of continuum model eliminates the need for additional Gaussian absorption or emission component around 10\,keV, which is often present in the spectra of accreting pulsars exhibiting cyclotron lines \citep{Coburn2001, Manikantan2023}. While we do note some residuals around 10 keV in the spectral fit of NuSTAR Obs II, this is not seen in other NuSTAR spectra. Therefore, we did not include this feature in our spectral model, to facilitate comparison of these spectral fits to the spectral modeling results of other RXTE, Suzaku and INTEGRAL observations which also did not include an additional model for this `10 keV' feature. Furthermore, the partial-covering absorption model provides a more physically consistent description of the NICER soft X-ray data than a blackbody component which was used previously in \cite{Sharma2023} for fitting the soft X-ray excess seen in the NuSTAR X-ray spectra. The line energy of the fundamental cyclotron line estimated from joint NICER and NuSTAR fits are consistent within error bars with the values estimated by NuSTAR-only fits for Obs III and Obs IV by \cite{Tamang2023}, although the spectral model used by them had an additional Gaussian absorption component to model the residuals around 8\,keV. \cite{Sharma2023} claimed the presence of two cyclotron lines at $\sim$ 22\,keV and $\sim$ 27\,keV, while fitting NuSTAR-only spectra for Obs III and IV. We do not find any evidence of second cyclotron line at 27\,keV in our spectral fits to the joint NICER and NuSTAR data. 
The other NuSTAR-only and NICER-only spectra were modeled with spectral models, described in Section \ref{sec:spectral analysis}, to obtain the spectral parameters as a function of orbital and superorbital phases. We also used results from previous RXTE, INTEGRAL and Suzaku observations from \cite{Hemphill2016} which had used the same {\tt mplcut} continuum model, to remove measurement biases in cyclotron line values and accurately estimate the temporal evolution of cyclotron line parameters. 
\par

\begin{figure}
    \centering
    \includegraphics[width=0.95\linewidth]{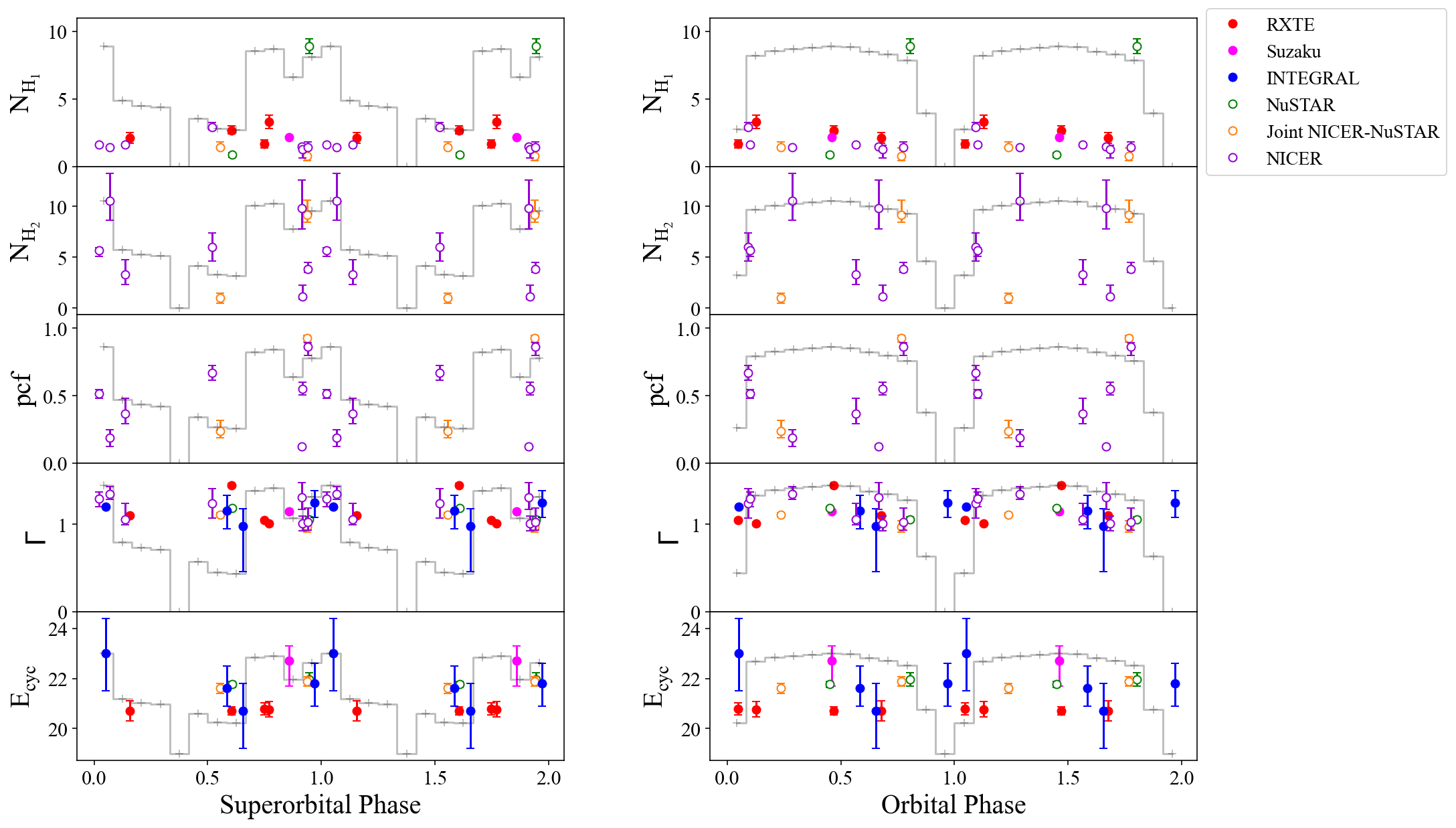}
    \caption{Spectral parameters of 4U 1538--52: $N_{H_{1}}$ ($10^{22}$ cm$^{-2})$, $N_{H_{2}}$ ($10^{22}$ cm$^{-2}$), pcf, photon index ($\Gamma$), and cyclotron line energy ($E_{cyc}$ in keV), plotted as a function of their superorbital (left) and orbital (right) phases. The gray histograms indicate the orbital and superorbital intensity profiles constructed with Swift--BAT and normalized for clarity. The spectral parameters estimated by fitting NICER and NuSTAR are shown as unfilled circles, archival values are shown as solid circles.}
    \label{fig:prev obs orbital + superorbital wINTEGRAL}
\end{figure}

Figure \ref{fig:prev obs orbital + superorbital wINTEGRAL} shows the spectral parameters of fully covering photoelectric absorption $N_{H_{1}}$, partial covering absorption $N_{H_{2}}$, partial covering fraction (pcf), photon-index $\Gamma$, and the cyclotron line energies E$_{cyc}$ as a function of superorbital and orbital phase, for all the NICER and NuSTAR observations (open circles) and archival RXTE, INTEGRAL and Suzaku observations (filled circles). The superorbital and orbital intensity profiles are plotted as gray histogram after normalizing them to the values of the spectral parameters. We do not find any obvious dependence of the spectral parameters on the superorbital and orbital phases covered by the pointed X-ray observations. Using a sine model, we provide upper limits to a sinusoidal modulation for N$_{H_{1}}$ to $23\%$ (superorbital) and $43\%$ (orbital), N$_{H_{2}}$ to $32\%$ (superorbital) and $6$\% (orbital), pcf to  $44\%$ (superorbital) and $98\%$ (orbital), $\Gamma$ of $7\%$ (superorbital) and $9\%$ (orbital) and E$_{cyc}$ to $3\%$ (superorbital) and $0.6\%$ (orbital). A significant source of uncertainty in spectral measurements derived from pointed X-ray observations arises from the challenge of disentangling short-term intensity and spectral variations caused by clumpy stellar winds in sgHMXBs from longer-term variations associated with large-scale density structures, such as CIRs \citep{Romano2024, Romano2025}. A detailed investigation of this issue will be presented in a forthcoming study based on long-term Swift-BAT X-ray Telescope monitoring of this source. 
\par
\begin{figure}
    \centering
\includegraphics[width=0.6\linewidth]{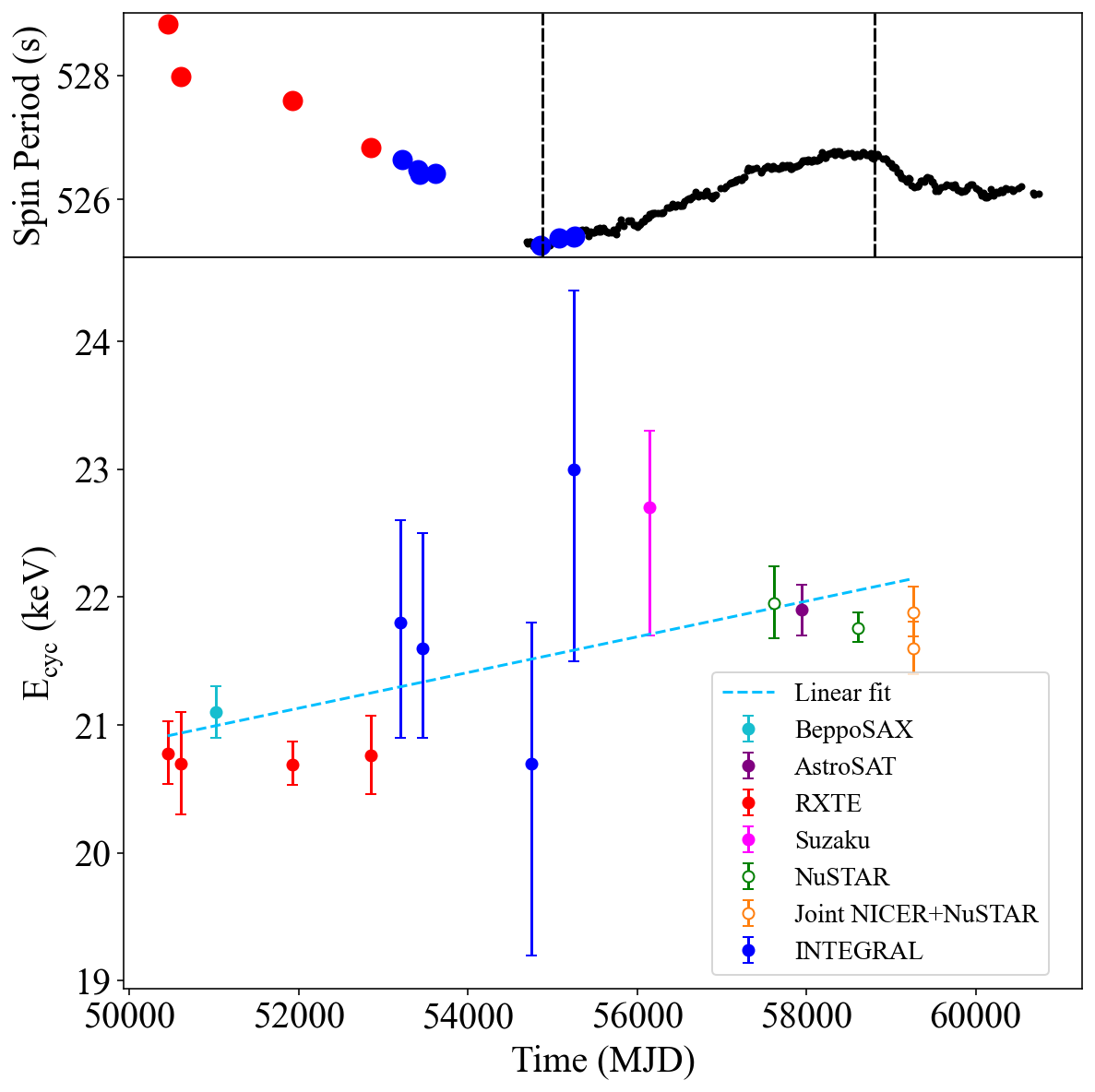}
    \caption{Top: The pulse period of 4U 1538--52 from Fermi-GMB (black), RXTE (red), and INTEGRAL (blue) observations, with dashed lines marking the dates of torque reversals. Bottom: Observations of 4U 1538--52's cyclotron line values as a function of MJD. The E$_{cyc}$ values estimated in this work are shown as unfilled circles, archival values are shown as solid circles. The blue dotted line is the constant linear fit, with a value of $0.051 \pm 0.022$\,keV yr$^{-1}$.}
    \label{fig:time Ecyc}
\end{figure}

The evolution of the cyclotron line energy and the pulse period over time is shown in Figure \ref{fig:time Ecyc}. We utilize the results from RXTE, Suzaku, INTEGRAL observations \citep{Hemphill2016} as well as the NuSTAR and joint NICER and NuSTAR observations analyzed in this work. We also include the results from a BeppoSaX observation \citep{Robba2001} and AstroSat \citep{Varun2019} but do not use them to create the linear fit, since the spectral modeling is different from the models used for RXTE, INTEGRAL, Suzaku results and the results used in this work.
The previous estimate by \cite{Hemphill2016} suggested an increase in the E$_{cyc}$ of 1.5\,keV over the 8.5 years between RXTE and Suzaku observations. \citep{Tamang2023} fit several functions: a stepwise constant, a broken linear fit and a linear fit, and reported a decrease in the cyclotron line energies after the Suzaku observation. We fit a linear model to the E$_{cyc}$ as a function of time and find an increase in the cyclotron line energies of $0.051 \pm 0.022$\,yr$^{-1}$, which is lower than reported by \cite{Hemphill2016}. Comparing the pulse period changes and the E$_{cyc}$ energies, we notice that there is a change in the trend following the first torque reversal around MJD 56000. However due to sparse coverage of the pointed observations in various epoch of pulse period changes, it is difficult to confirm this trend.

\subsection{Comparison with the Models}
Various models invoking precessing accretion disks have not been able to account for the presence of the second harmonic component of the superorbital modulation in the power spectrum \citep{Islam2023}. Alternatively, tidal oscillations in nonsynchronously rotating supergiant stars can give rise to structured stellar winds, which may drive the observed superorbital modulations in sgHMXBs \citep{koenigsberger2006}. In this model, superorbital modulation would be caused by tidal interactions which produce perturbations on the external layers of the stars and in turn modulate the companion star's stellar wind. However, in eccentric systems such as 4U 1538--52, the modulation is observed near the binary periastron and is therefore likely associated with the orbital period \citep{Moreno2011}.
\par
CIRs are long-lived, large-scale spiral-shaped density and velocity structures that develop within the stellar winds of massive stars. Initially identified in the solar wind, CIRs have since been detected around other stars through the observation of discrete absorption components (DACs) in optical  spectra. These structures are believed to originate from surface irregularities such as dark or bright spots, magnetic loops, or nonradial pulsations, which induce spiral-shaped perturbations in the wind density and velocity extending up to several tens of stellar radii \citep{CranmerOwocki1996}.
Models incorporating CIRs in the stellar winds of supergiant stars can account for the presence of both the fundamental and harmonic components of superorbital modulations observed in the power spectra of sgHMXBs \citep{Bozzo2017}. In this framework, the superorbital modulations correspond to the beat frequency between the CIR and the NS orbital period around the supergiant star. The orbital period of the CIR can be either longer ("long CIR period") or shorter ("short CIR period") than the orbital period. \cite{corbet2021} found that the short CIR periods lie between 0.7--0.83 times the orbital period, and long CIR periods between 1.33--1.60 times the orbital period for the six sgHMXBs exhibiting superorbital modulation.\citet{corbet2021} proposed two possible mechanisms that could yield a CIR period close to, but not identical with, the orbital period: (i) in eccentric systems, tidal synchronization near periastron may alter the primary star’s rotation period; and (ii) differential rotation of the supergiant may cause CIRs to form in regions rotating faster than the star’s mean rotational period.
\par
Pulse period changes from torque reversals have been observed in several sgHMXBs, such as 4U 1538–-52 and Vela X-–1, which are not directly related with the presence of accretion disks in these systems. Long-term variations in the properties of the stellar winds are proposed as the primary cause of these torque reversals, suggesting the influence of large-scale structures such as CIRs within the stellar winds. A possible link between superorbital modulations and spin-period evolution has been noted in 2S 0114+650 and 4U 1538–52, both of which exhibit superorbital variability. However, for many other systems, the absence of continuous spin-period monitoring hampers efforts to explore potential correlations between spin evolution and superorbital modulation. Limited pointed X-ray observations of 4U 1909+07 suggest that variations in the amplitude of the superorbital modulation may be associated with changes in the neutron star’s pulse period \citep{Islam2023}. In contrast, for a system such as Vela X–1, where torque reversals are attributed to structured stellar winds, no significant superorbital modulation is detected in its Swift-BAT lightcurve. A possible interpretation is that torque reversals in Vela X--1 take place on shorter timescales than those measured for 4U 1538--52. This would suggest that the formation and dissipation of large-scale wind structures in the stellar wind of HD 77581 (the supergiant companion in Vela X–1) occur more rapidly than in the wind of QV Nor (the supergiant companion in 4U 1538–52). Alternatively, if CIRs are present in the stellar wind of HD 77581, their characteristic timescales or locations may fall outside the range required to produce detectable superorbital modulation in Vela X--1. Optical observations of large scale structures in the stellar winds of the supergiant stars would be crucial in understanding their formation and dissipation mechanisms.

\section{Summary}
\label{sec:Summary}
The sgHMXB 4U 1538--52 has a pulse-period history which shows at least three epochs of pulse period changes uncorrelated with the X-ray flux and possibly caused by variations in the stellar wind, transient disk formation, or tidal interactions. This study examines the relationship between 4U 1538--52’s superorbital modulations, spectral and timing properties, and torque reversals using long-term monitoring data from Fermi-GBM, Swift-BAT, and MAXI-GSC, along with pointed observations from NICER, NuSTAR, and other missions. The pulse period history of 4U 1538--52 is characterized by changing pulse period behavior as Spin Epoch I (MJD 53415 to 54880, with the pulsar in spin-up phase), Spin Epoch II (MJD 54880 to MJD 58400, marked by a torque reversal at MJD 54880 which changed the behavior from spin-up to spin-down) and Spin Epoch III  (MJD 58400 to MJD 60750), with the pulsar having an erratic sinusoidal spin change that tends towards spin-up,\citep{Sharma2023}.
\par
Superorbital modulations are detected in MAXI-GSC 2--20\,keV lightcurves, along with Swift-BAT 15--50\,keV lightcurves with a period of 14.9130$\pm$0.0026\,d. The dynamic power spectrum constructed with Swift-BAT lightcurves shows the fundamental frequency of the superorbital modulations present throughout, apart from the time interval of MJD 57650 to MJD 58800 during which the fundamental frequency becomes weaker and the second harmonic is stronger. We only see the fundamental frequency of the superorbital period in MAXI-GSC lightcurves. The shape of the superorbital intensity profiles in the time segments, where the fundamental frequency or the second harmonic of the superorbital modulation is present or in the segments corresponding to change in the spin behavior, show a similar single peaked profile with a shift in the phase of the maximum X-ray flux. 
\par
The energy resolved pulse profiles constructed from NICER and NuSTAR observations show no change in shape at different orbital, superorbital and epochs of spin changes. Similarly, the spectral parameters like photoelectric absorption $N_{H}$, photon-index $\Gamma$ and line energies of the cyclotron line E$_{cyc}$ show no obvious dependence on either the superorbital or orbital phase. However, it is difficult to disentangle the contribution of short term variability in these sgHMXBs from long-term variations in spectral parameters expected from mechanisms causing these superorbital modulations and we leave this issue for a future study. 
\par
We also investigate the long-term evolution of the cyclotron line energies and find that there is a linear increase in the line energies at the rate of 0.051$\pm$0.022\,keV yr$^{-1}$, smaller than that estimated by \cite{Hemphill2016}. Figure \ref{fig:time Ecyc} shows cyclotron line energy as a function of time and pulse period of 4U 1538--52. We notice a change in the trend of increase in the cyclotron line energy around the epoch of torque reversal, however long-term X-ray observations are required to confirm this trend. 
\par
The presence of large scale structures like CIRs in the stellar wind of supergiant stars has been invoked to explain the superorbital modulations in sgHMXBs. A possible link between the superorbital modulations and pulse period changes have been observed in 2S 0114+650, although they have been attributed to transient accretion disks or tidal oscillations \citep{Hu2017, koenigsberger2006}. These large scale structures can also explain the torque reversals seen in 4U 1538--52 and Vela X--1, although superorbital modulation is not detected in Vela X--1, which provides constraints on the formation and rotational velocities of these structures. Regular monitoring of these sources are required with Fermi-GBM, and sensitive all-sky monitors such as the enhanced X-ray Timing and Polarimetry mission \citep[eXTP; ][]{zhang2016} are required to further understand the connection between superorbital modulations and pulse period changes. 

\begin{acknowledgements}
We thank the anonymous referee for the constructive
comments that helped improve the manuscript.
We thank members of the XMAG collaboration for useful discussions. This research made use of data from the NuSTAR mission, a project led by the California Institute of Technology, managed by the Jet Propulsion Laboratory, and funded by the National Aeronautics and Space Administration. Data analysis was performed using the NuSTAR Data Analysis Software (NuSTARDAS), jointly developed by the ASI Science Data Center (SSDC, Italy) and the California Institute of Technology (USA). This research has made use of data and/or software provided by the High Energy Astrophysics Science Archive Research Center (HEASARC), which is a service of the Astrophysics Science Division at NASA/GSFC. This research uses Swift-BAT transient monitor results provided by the Swift-BAT team. The scientific results reported here are based on observations made by the NICER X-ray observatory. This research has made use of MAXI data provided by RIKEN, JAXA and the MAXI team. We thank the Fermi-GBM team for providing
the public data used in this analysis. Support for this work was provided by NASA through grant No. 80NSSC25K7622 and 80NSSC21K0022. The work was partially supported by NASA under award numbers 80GSFC21M0006. \

\software{XSPEC (v12.15.0d; \citealt{Arnaud1996}); Stingray (v1.1.2; \citealt{2019ApJ...881...39H, Huppenkothen2019, Stingray}); Astropy (v6.1.2; \citealt{Astropy2022}; Robot (v4.94; \citealt{Corbet1992}))
}

\end{acknowledgements}


\end{document}